\documentclass[a4paper,11pt]{article}
\usepackage{jcappub} 
\usepackage{fancyhdr}
\usepackage{graphicx}
\usepackage{wrapfig}
\usepackage{times}
\usepackage{color}
\usepackage{hyperref}
\usepackage{amsmath}
\usepackage{tabularx} 
\usepackage{multirow} 
\usepackage{enumitem}
\usepackage{orcidlink}

\title{Non-linear reconstruction of general dark energy theories}

\author[a, b]{Yunhao Gao~\orcidlink{0009-0000-6958-1139},}
\author[c]{Baojiu Li~\orcidlink{0000-0002-1098-9188},}
\author[a, b]{and Jie Wang~\orcidlink{0000-0002-9937-2351}}
\affiliation[a]{National Astronomical Observatories, Chinese Academy of Sciences, Beijing, 100101, China}
\affiliation[b]{School of Astronomy and Space Sciences, University of Chinese Academy of Sciences, Beijing, 100049, China}
\affiliation[c]{Institute for Computational Cosmology, Department of Physics, Durham University, Durham DH1 3LE, UK}

\emailAdd{gaoyh@bao.ac.cn} 
\emailAdd{baojiu.li@durham.ac.uk}
\emailAdd{jie.wang@nao.cas.cn}
\abstract{The large variety and number of dark energy (DE) theories make it impractical to perform detailed analyses on a case-by-case basis, which has motivated proposals to ``parameterize" theories to reduce the size of theory space. The leading approach to do this is the effective field theory of dark energy (EFTofDE), which can describe general Horndeski-type theories with a small number of observationally accessible time-dependent functions. However, the EFTofDE primarily works for linear perturbations, and extending it to obtain a fully non-linear description of DE theories, which is critical for theories with screening mechanisms, is challenging. In this paper, we present a general method for reconstructing the non-linear DE Lagrangian from the background expansion history and certain linear-perturbation quantities, building upon the EFTofDE framework. Using numerical examples, we demonstrate that this method is applicable to a wide range of single-scalar-field dark energy and modified gravity theories, including quintessence, scalar-tensor theory, $k$-essence, and generalized cubic Galileon with shift symmetry. For each of these theories, we discuss the validity of the method and factors affecting its results. While this method involves solving differential equations, we find that the initial conditions are not important for quintessence, scalar-tensor theory and $k$-essence, while for shift-symmetric cubic Galileon, the generic tracker solution can help transform differential equations into algebraic equations. This offers a useful framework to connect cosmological observations at the background and linear-perturbation levels to the underlying non-linear dynamics of dark energy, and will enable cosmological simulations to analyze and examine DE theories systematically and in much greater detail.}

\begin{document}
\maketitle
\flushbottom

\section{Introduction}
\label{sec:intro}

The origin of the late-time accelerated expansion of the Universe is one of the most important open questions in cosmology. Since the early discoveries of the cosmic acceleration \cite{dark_energy_supernovae_Riess,dark_energy_supernovae_Perlmutter}, a large number of dark energy (DE) and modified gravity (MG) models, which attempt to explain this phenomenon, have been proposed in the literature, the notable examples of which include quintessence \cite{quintessence_theory1, quintessence_theory2, quintessence_theory3, quintessence_theory4, quintessence_theory5}, $f(R)$ gravity \cite{FR_theory1, FR_theory2}, $k$-essence \cite{kessence_theory1, kessence_theory2}, Horndeski theory (or generalized Galileon) \cite{Horndeski_theory, GG_theory} and the Dvali-Gabadadze-Porrati (DGP) model \cite{DGP_theory}. 

Cosmological observables, such as baryonic acoustic oscillation (BAO), luminosity distance of type Ia supernovae (SNIe), gravitational lensing and cosmic microwave background (CMB), enable us to measure the background expansion and the structure formation of our universe with high precision, offering the potential of examining different DE theories and excluding non-viable ones (see, e.g., \cite{probes_review1, probes_review2} for reviews of these probes). Nevertheless, the diversity and huge number of these theories make individual evaluation both methodologically cumbersome and computationally expensive.

A common method to tackle the issues and difficulties in this regard is ``parametrization", which extracts common characters of various dark energy theories and describes them using a few parameters. A primary example is the Chevallier-Polarski-Linder (CPL), or $w_0$-$w_a$ parametrization \cite{CPL1, CPL2}, which assumes the evolution of the equation of state of dark energy, $w_{\textrm{DE}}$, is a linear function of the scale factor $a$. This is a purely phenomenological approach, with no consideration of the microscopic physics. In the area of modified gravity, some time- and scale-dependent factors are popularly utilized to measure the deviation from general relativity \cite{MG_parametrize1, MG_parametrize2}. Apart from these formalisms, there are some more general parameterizations, like the effective field theory of dark energy (EFTofDE), which characterizes a broad class of single-scalar-field dark energy models through a Lagrangian involving several time-dependent functions and spacetime perturbations up to the second order \cite{EFT_theory1, EFT_theory2, EFT_theory3, EFT_theory4, EFT_review}. A nice thing about the EFTofDE is that it is mathematically and physically motivated so that, while it is still phenomenological, there is a closer connection to rigorous fundamental Lagrangians. 

Despite the simplicity and the success achieved by these parameterization schemes, the majority of them concern only the background and linear perturbation evolution of dark energy theories, while a fully self-consistent description of non-linear phenomena with relatively few parameters remains a challenge. For example, Refs.~\cite{nonlinear_EFT1} and \cite{nonlinear_EFT2} present some interesting attempts to extend the EFTofDE formalism to the quasi-non-linear scales, by including terms beyond the usual second order (at the third and fourth orders) in the action. However, as we will explain in Section \ref{sec:overview}, these extensions would in general lead to a more complicated DE Lagrangian, and that will get worse if one aims to go to more non-linear scales by including more terms following this approach. This prevents us from comparing the non-linear properties of different theories through the parametrization.



Another viable strategy to circumvent the problem of describing non-linear dynamics is to reconstruct the functional form of the dark energy Lagrangians through the background and linear observations. Similar methodologies have been extensively researched in the literature, including reconstructing quintessence \cite{quintessence_reconstruct1, quintessence_reconstruct2, quintessence_reconstruct3, quintessence_reconstruct4, quintessence_reconstruct5, quintessence_reconstruct6, quintessence_reconstruct7, STT_reconstruct1, STT_reconstruct2, STT_reconstruct3}, scalar-tensor theories \cite{STT_reconstruct7, STT_reconstruct4, STT_reconstruct5, STT_reconstruct6, STT_reconstruct8, MG_tomography}, $f(R)$, $f(R, T)$, $f(R, G)$, and $f(T)$ theories \cite{FR_reconstruct1, FR_reconstruct2, FR_reconstruct3, FR_reconstruct4, FR_reconstruct5, FR_reconstruct6, FR_reconstruct7, FR_reconstruct8, FR_reconstruct9, FR_reconstruct10}, $k$-essence \cite{kessence_reconstruct1, kessence_reconstruct2, kessence_reconstruct3, kessence_reconstruct4}, ghost condensation \cite{ghost_condensation_reconstruct1, ghost_condensation_reconstruct2}, generalized cubic Galileon\cite{cG_reconstruct1, cG_reconstruct2, cG_reconstruct3, cG_reconstruct4, cG_reconstruct5, cG_reconstruct6, cG_reconstruct7}, and even models connected to the inflation, such as $k$-inflation \cite{k_inflation1, k_inflation2, k_inflation3} and G-inflation \cite{G_inflation1}. In spite of the broad theoretical coverage, most of these reconstruction works are restricted to one specific DE theory that is described by pre-chosen functions in the Lagrangian. These approaches have the benefit of being targeted at the price of giving up generality. 

The first approach described above, which add additional terms to the parameterization to produce non-linearity in the Lagrangian, is based on the idea that the background and linear-perturbation parameters can, in some sense, be considered as the first few coefficients of some ``Taylor expansion of the full theory". In an ``infinite" theory space, i.e., one where we could include arbitrary terms into the parameterized Lagrangian without causing inconsistencies, this is a natural thing to do. The price to pay, as mentioned above, is evidently that the resulting DE theory can be ``infinitely'' complicated. Fortunately, General Relativity is a very ``rigid" theory which can only be modified in a finite number of ways if it is to remain healthy and consistent, and if certain physics principles are to be respected (e.g., the Ostrogradsky theorem). Usually, it is the same terms in the Lagrangian that dictate both the background/linear and fully non-linear dynamics of the theory, and so the latter must be ``correlated" with each other somehow. By exploiting such correlations, it may be possible to infer the fully non-linear dynamics from the background evolution (and a few linear-perturbation properties if needed) of a theory. Apparently, this approach would be easier in practice if there are relatively fewer terms in the parameterized Lagrangian, so its implementation may still be challenging if we insist on having the ``most general possible" DE and MG theories\footnote{For this reason the examples we give in this paper will be restricted to certain generic classes of theories}. However, \textcolor{black}{when investigating DE theories with specific physical properties, in practice the favored DE models are typically those simple enough and with few enough additional degrees of freedom, and} we feel that there is not a strong desire to have overly-complicated DE theories that include multiple new physics, and for this reason there must be a middle way to have parameterizations for limited but still generic classes of theories, that are simple (i.e., with relatively few parameters) but allow to ``reconstruct" the full non-linear theory.

From a theoretical viewpoint, being able to reconstruct the full theory from a simple parametrization is appealing since it makes it easier to perform comprehensive analyses of generic DE theories. From an observational perspective, new precision data (such as the recent DESI analysis \cite{DESI:2025zgx}) have increasingly allowed us to measure the expansion history of the Universe, and these measurements should naturally be fed into the study of DE theories, making the latter more targeted and dedicated.

In this paper, we employ the effective field theory of dark energy to propose a new reconstruction methodology of this flavor. We note here that Refs.~\cite{EFT_reconstruct1, EFT_reconstruct2, EFT_reconstruct3, EFT_reconstruct4} have introduced an EFTofDE-based reconstruction method for ``covariant Horndeski theories", and will explain in the following sections how our method differs. As we will demonstrate in Section \ref{sec:overview}, in the EFTofDE framework, the linear evolution of the dark energy is completely encoded in several time-dependent functions, which can be constrained by background and linear-perturbation observations. As mentioned above, for every specific class of dark energy theory involving a scalar field, its Lagrangian can be connected to these EFT functions. From these connections, the time evolution of both the scalar field and the functions in the DE Lagrangian can be determined, and subsequently, the Lagrangian's dependence on the scalar field and its derivatives can also be derived.

In the following part of this paper, we first briefly overview some concepts about the Horndeski gravity and the EFTofDE, as well as the basic idea of our reconstruction framework in Section \ref{sec:overview}. In the next three sections, we implement this method on different dark energy models, i.e., quintessence and scalar tensor theory in Section \ref{sec:quintessence}, $k$-essence in Section \ref{sec:kessence}, and cubic Galileon with shift symmetry in Section \ref{sec:galileon}. We then conclude this paper and prospect the applications of our method in Section \ref{sec:discuss}.

\section{An overview of Horndeski gravity and the EFTofDE}
\label{sec:overview}

The action of Horndeski gravity, or generalized Galileon, can be written as \footnote{In the literature, there are several definitions of the action of generalized Galileon, slightly different from each other. In this paper, we choose the same notation convention as Eq. (50)--(53) of \cite{EFT_theory3}.}
\begin{equation}
    S_{\textrm{H}}=\int \textrm{d}^4x\sqrt{-g}\left(L_2+L_3+L_4+L_5\right),
\end{equation}
with $L_2$ -- $L_5$ given as
\begin{equation}
\begin{aligned}
    L_2&=K(\phi,X),\\
    L_3&=G_3(\phi,X)\square\phi,\\
    L_4&=G_4(\phi,X)R-2G_{4X}(\phi,X)\left[(\square\phi)^2-\phi_{;\mu\nu}\phi^{;\mu\nu}\right],\\
    L_5&=G_5(\phi,X)G_{\mu\nu}\phi^{;\mu\nu}+\frac{1}{3}G_{5X}(\phi,X)\left[(\square\phi)^3-3\square\phi\phi_{;\mu\nu}\phi^{;\mu\nu}+2\phi_{;\mu\nu}\phi^{;\mu\sigma}\phi^{;\nu}_{;\sigma}\right],
\end{aligned}
\end{equation}
where $X=g^{\mu\nu}\partial_\mu\phi\partial_\nu\phi$,  $\square\phi=g^{\mu\nu}\nabla_\mu\nabla_\nu\phi$, and a semicolon denotes the covariant derivative, e.g., $\phi_{;\mu}=\nabla_\mu\phi$. Hereafter, subscript $_\phi$ and $_X$ denotes derivatives w.r.t. $\phi$ and $X$. The Horndeski theory, which includes a number of interesting dark energy and modified gravity models, is the most general 4-dimensional scalar field theory with no higher than second order derivatives in the field equation, whose implications to the late-time acceleration of the Universe have been extensively researched in the literature (see \cite{Horndeski_review} for a review). The nearly simultaneous detections of gravitational waves and their electromagnetic counterparts in 2017 \cite{LIGOScientific:2017vwq} confirm that there can be very little, if any, difference between the propagation speed of tensor perturbations and the speed of light, which effectively places the constraint that $G_5=0$ and $G_4$ be independent of $X$ \cite{GW17_1, GW17_2, GW17_3, GW17_4}. Thus, the dark energy models that we will consider are a sub-class of generalized Galileon theory whose action can be written as
\begin{equation}\label{GG action}
    S_{\textrm{DE}}=\int \textrm{d}^4x\sqrt{-g}\left[\frac{M^2_*}{2}G_4(\phi)R+K(\phi, X)+G_3(\phi, X)\square\phi\right],
\end{equation}
where $M_*=1/\sqrt{8\pi G}$ is the so-called reduced Plank mass.

The characteristics of different subclasses of the Horndeski theory on the background and linear scales can be captured by the EFTofDE parametrization \cite{EFT_theory1, EFT_theory2, EFT_theory3, EFT_theory4}. Under the unitary gauge, where the spatial perturbation of dark energy vanishes, the action can be written as
\begin{equation}\label{EFT action}
    S_{\textrm{EFT}}=\int \textrm{d}^4x\sqrt{-g}\bigg[\frac{M^2_*}{2}f(t)R-\Lambda(t)-c(t)g^{00}+\frac{M^4_2(t)}{2}(\delta g^{00})^2-\frac{\bar{M}^3_1(t)}{2}\delta g^{00}\delta K+\cdots\bigg],
\end{equation}
where we call time functions \{$f(t)$, $\Lambda(t)$, $c(t)$, $M^4_2(t)$, $\bar{M}^3_1(t)$, $\cdots$\} ``EFT parameters", and $\delta g^{00}$ and $\delta K$ are the perturbation of the upper time-time component of the metric and the trace of the extrinsic curvature, respectively. From \cite{EFT_theory4}, the relation between the EFT parameters and $K$, $G_3$, and $G_4$ in the action Eq.~\eqref{GG action} can be specified as
\begin{equation}\label{EFT relation}
\begin{aligned}
    f(t)=&G_4(\phi),\\
    \Lambda(t)=&\dot{\phi}^2G_{3X}(\phi, X)(3H\dot{\phi}+\ddot{\phi})-K_X(\phi, X)\dot{\phi}^2-K(\phi, X),\\
    c(t)=&\dot{\phi}^2G_{3X}(\phi, X)(3H\dot{\phi}-\ddot{\phi})+\dot{\phi}^2G_{3\phi}(\phi, X)-K_X(\phi, X)\dot{\phi}^2,\\
    M^4_2(t)=&\frac{1}{2}\dot{\phi}^2G_{3X}(\phi, X)(3H\dot{\phi}+\ddot{\phi})-3H\dot{\phi}^5G_{3XX}(\phi, X)-\frac{1}{2}\dot{\phi}^4G_{3\phi X}(\phi, X)\\&+K_{XX}(\phi, X)\dot{\phi}^4,\\
    \bar{M}^3_1(t)=&-2\dot{\phi}^3G_{3X}(\phi, X),
\end{aligned}
\end{equation}
where $H$ is the Hubble parameter, and we have used the background values of $\phi$ and $X$.\footnote{For the sake of simplicity, we use the same symbols $\phi$ and $X$ to denote both the background quantities, $\phi(t)$, $X(t) = -\dot{\phi}^2$, and the full spacetime-dependent fields, $\phi(x^\mu)$ and $X(x^\mu) = g^{\mu\nu}\partial_\mu\phi\partial_\nu\phi$. The distinction between background and full quantities should be clear from the context, and we believe that this slight abuse of notation will not cause confusion.} Throughout this paper, the upper dot denotes the derivative with respect to the cosmic time $t$. On the background, the first three parameters are related as \cite{EFT_review}
\begin{equation}\label{background EFT}
\begin{aligned}
    c&=\frac{1}{2}\left(\rho_{\textrm{DE}}+P_{\textrm{DE}}\right)+\frac{1}{2}\left(-\ddot{f}+H\dot{f}\right)M_*^2,\\
    \Lambda&=\frac{1}{2}\left(\rho_{\textrm{DE}}-P_{\textrm{DE}}\right)+\frac{1}{2}\left(\ddot{f}+5H\dot{f}\right)M_*^2,
\end{aligned}
\end{equation}
where $\rho_{\textrm{DE}}$ and $P_{\textrm{DE}}$ are the density and pressure of dark energy respectively. If there is no coupling between the scalar field and gravity, i.e., $f=1$, by using Eq.~\eqref{background EFT}, the conservation equation for dark energy, $\dot{\rho}_{\textrm{DE}}+3H(\rho_{\textrm{DE}}+P_{\textrm{DE}})=0$, is
\begin{equation}\label{conservation equation}
    \dot{c}+\dot{\Lambda}+6Hc=0.
\end{equation}
For the physical properties of linear perturbations, the (first five) EFT parameters can be connected with the alpha-basis \cite{alpha_basis},
\begin{equation}\label{alpha_basis}
    \alpha_K=\frac{2c+4M_2^4}{M^2_*H^2},\quad \alpha_B=-\frac{\bar{M}_1^3}{M^2_*H},\quad \alpha_M=\frac{\dot{f}}{Hf},
\end{equation}
where $\alpha_K$ is linked to the kinetic energy of the scalar perturbations, $\alpha_B$ encodes the strength of the mixing between the kinetic terms of the scalar field and metric, and $\alpha_M$ measures the changing rate of the effective Planck mass. We note that there are well-established Boltzmann codes, e.g., \texttt{EFTCAMB} \cite{EFTCAMB1, EFTCAMB2} and \texttt{hi\_class} \cite{hiclass1, hiclass2}, which are able to explore the influences of the EFT parameters on observables in the linear perturbation regime. 

Despite the wide application of EFTofDE in background and linear-perturbation evolutions, and certain attempts to extend its action beyond the linear scales \cite{nonlinear_EFT1, nonlinear_EFT2}, this formalism is inconvenient and unnatural for studying the non-linear evolution of a specific DE model. To elucidate this problem, we convert the $k$-essence model, where $G_3=0$ and $G_4=1$, into the EFT parametrization as an example. Its action is
\begin{equation}\label{k essence action}
    S=\int{\rm d}^4x\sqrt{-g}\left[\frac{M^2_*}{2}R+K(\phi, X)\right].
\end{equation}
In the unitary gauge, we have $\phi(t, x^i)=\phi(t)$, which is already a background quantity, and
\begin{equation}
    X\equiv g^{\mu\nu}\partial_\mu\phi\partial_\nu\phi=g^{00} \dot{\phi}^2=\left(-1+\delta g^{00}\right)\dot{\phi}^2.
\end{equation}
Expand the function $K$ w.r.t. $X$ at the point $-\dot{\phi}^2(t)$, the action of Eq.~\eqref{k essence action} becomes
\begin{equation}
\begin{aligned}
    S=&\int{\rm d}^4x\sqrt{-g}\bigg[\frac{M^2_*}{2}R+K(\phi, -\dot{\phi}^2)+K_X\left(\phi, -\dot{\phi}^2\right)\dot{\phi}^2\delta g^{00}\\&+\frac{1}{2}K_{XX}\left(\phi, -\dot{\phi}^2\right)\dot{\phi}^4\left(\delta g^{00}\right)^2+\cdots+\frac{1}{n!}\frac{\partial^n K}{\partial X^n}\left(\phi, -\dot{\phi}^2\right)\dot{\phi}^{2n}\left(\delta g^{00}\right)^n+\cdots\bigg].
\end{aligned}
\end{equation}
Now, the EFT action can be written as
\begin{equation}\label{non-linear EFT}
\begin{aligned}
    S_{\textrm{EFT}}=\int \textrm{d}^4x\sqrt{-g}\bigg[&\frac{M^2_*}{2}R-\Lambda(t)-c(t)g^{00}+\frac{M^4_2(t)}{2}\left(\delta g^{00}\right)^2\\&-\frac{M^4_3(t)}{3!}\left(\delta g^{00}\right)^3+\cdots+(-1)^n\frac{M^4_n(t)}{n!}\left(\delta g^{00}\right)^n+\cdots\bigg],
\end{aligned}
\end{equation}
with
\begin{equation}
\begin{aligned}
    &f(t)=1,\quad \Lambda(t)=K_X(\phi, X)X-K(\phi, X),\quad c(t)=K_X(\phi, X)X,\\&M^4_2(t)=\frac{\partial^2 K}{\partial X^2}(\phi, X)X^2,\cdots,M^4_n(t)=\frac{\partial^n K}{\partial X^n}(\phi, X)X^n,\cdots
\end{aligned}
\end{equation}
where we have used the fact that $X(t)=-\dot{\phi}^2$ on the background. For all $n\geq3$, the terms involving $M^4_n(t)$ in Eq.~\eqref{non-linear EFT} correspond to the non-linear evolution. (These are part of the terms mentioned in \cite{nonlinear_EFT1, nonlinear_EFT2}, and Refs.~\cite{EFT_reconstruct1, EFT_reconstruct2, EFT_reconstruct3, EFT_reconstruct4} used these higher-order terms to obtain a more accurate reconstruction.) While adding more terms like that allows for increasingly accurate description of non-linear effects, this comes at the cost of drastically increasing the complexity of the effective action. Each additional term introduces new time-dependent coefficients that must be individually constrained, making the EFT framework increasingly unwieldy in practice. As a concrete example, if $K(\phi,X)=\sqrt{-X}$, the full action is simple, but the EFT expansion could lead to an infinite tower of terms, and hence parameters. Consequently, for non-linear scales, working directly with the original theory $K(\phi, X)$ often proves more efficient than its EFT counterpart.

As mentioned in Section \ref{sec:intro}, and as the above example shows, for a specific dark energy model, the first few terms in its corresponding EFT parametrization are not necessarily independent of the higher-order terms, and using the first 5 EFT parameters shown in Eq.~\eqref{EFT action} might already be sufficient to determine the $\phi$ and $X$ dependence of the functions $K$, $G_3$ and $G_4$ (or at least when the forms of $K$, $G_3$ and $G_4$ are simple enough). To achieve this, we note that in Eq.~(\ref{EFT relation}) we can think of $K$, $G_3$, and $G_4$ as functions of time, for example $G_3(t)=G_3(\phi(t), X(t))$. As a result, once we fix all the left-hand side of Eq.~(\ref{EFT relation}), these formulae can actually be treated as 5 ordinary differential equations for $\phi(t)$, $X(t)$, $K(t)$, $G_3(t)$ and $G_4(t)$. 

This thought plays an important role in our reconstruction framework to be introduced below. In the following three sections, we will present some concrete examples to show that, under some extra but still general assumptions for $K$, $G_3$, and $G_4$, we can fully reconstruct their dependence on $\phi$ and $X$ by solving for $\phi(t)$, $X(t)$, $K(t)$, $G_3(t)$ and $G_4(t)$, provided that we have fixed the time evolution of the EFT parameters from the observations. In this sense, the fully non-linear action of a dark energy theory may be reconstructed by knowing the linear theory or even the background evolution only.

\section{Reconstruction of quintessence and scalar-tensor theories}
\label{sec:quintessence}


We first illustrate the reconstruction method in the simplest DE model: quintessence. This model is specified as
\begin{equation}\label{quintessence}
    K=-\frac{1}{2}X-V(\phi),\quad G_3=0,\quad G_4=1.
\end{equation}
We note here again that reconstructions of quintessence, or the non-minimally coupled scalar-tensor theories, are not new and there have already been numerous works on this topic. For example, to get the form of $V(\phi)$, \cite{quintessence_reconstruct1, quintessence_reconstruct3} measured the redshift-dependent coordinate distance $r(z)$, and \cite{quintessence_reconstruct2, quintessence_reconstruct6} directly assumed the evolution of the dark energy density $\rho_{\textrm{DE}}(t)$. In Refs.~\cite{STT_reconstruct1, STT_reconstruct2, STT_reconstruct3, STT_reconstruct4, STT_reconstruct5, STT_reconstruct6}, the authors fixed the $\phi$-dependence of $t$ and $t$-dependence of the Hubble parameter to reconstruct the dilaton and scalar-Gauss-Bonnet gravity. Apart from these examples, for scalar-tensor theory, Ref.~\cite{STT_reconstruct7} use the evolution of the Hubble parameter and the matter perturbation to reconstruct the DE Lagrangian, and \cite{MG_tomography} proposed the idea of the modified gravity tomography, demonstrated that a series of scalar-tensor theories can be reconstructed through the mass of the scalar field $m(\phi(t))$ and the coupling strength $\beta(\phi(t))$. Additionally, Ref.~\cite{quintessence_reconstruct7} proposed an EFTofDE-based reconstruction method, very similar to the one we will derive in this section, for quintessence. Despite these existing methodologies in the literature, the reconstruction framework of these two models is conducive to the completeness of our discussion and serves as a conceptual bridge to our approach to the more complicated DE models. As a result, here we outline the reconstruction process of quintessence, as well as scalar-tensor theories, as the first application of our method.

\begin{figure}[t]
\begin{center}
\includegraphics[width=1\textwidth]{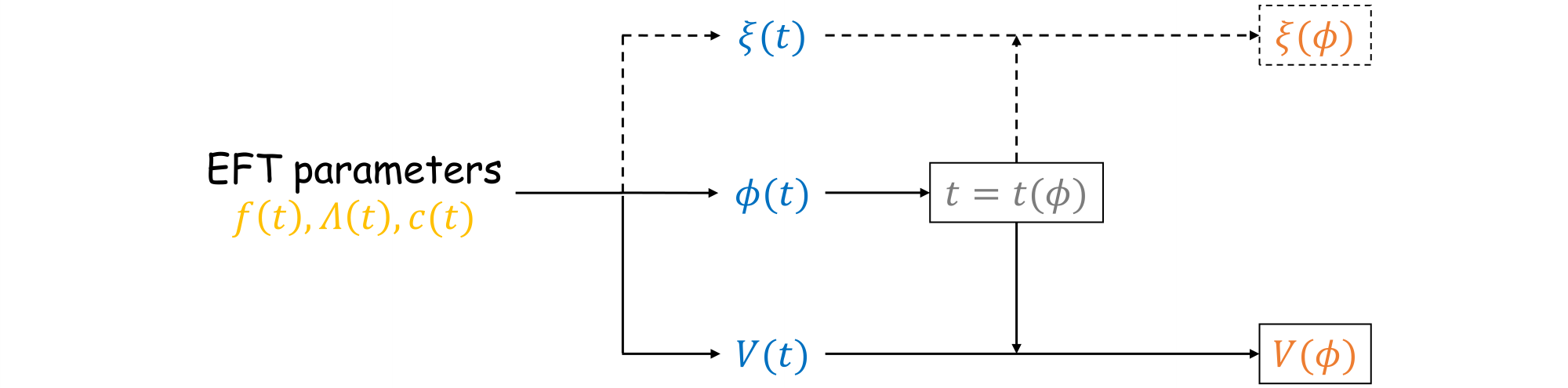}
\caption{A sketch of the reconstruction scheme for the quintessence (the solid line) and scalar tensor models (including the dashed line). The EFT parameters are shown in yellow. Quantities shown in blue are solved for from the differential equations, while quantities in orange are reconstructed.}
\label{fig:fc_quint}
\end{center}
\end{figure}

For quintessence, substituting the definition Eq.~\eqref{quintessence} into Eq.~\eqref{EFT relation}, we get
\begin{equation}
    c(t)=\frac{1}{2}\dot{\phi}^2,\quad \Lambda(t)=V\big(\phi(t)\big)=V(t),
\end{equation}
and $f(t)=1$, $M_2^4(t)=\bar{M}_1^3(t)=0$. If we fix $\Lambda(t)$ and $c(t)$, we automatically get the evolution of $V(t)$. If $\dot{\phi}$ is assumed to be positive all the time,\footnote{The sign of $\dot{\phi}$ is insignificant. For instance, for quintessence, when $\dot{\phi} < 0$, we can just transform $V$ into $-V$, and $\dot{\phi}$ into $-\dot{\phi}$, then the background evolution eq.~\eqref{quintessence background EoM} will be the same. Thus, in this paper, whenever we assume $\phi$ is monotonic, we will always stipulate that $\dot{\phi}>0$.} we thereby get $\phi(t)$ through an integration:
\begin{equation}\label{phi integral 1}
    \phi(t)=\int^t_{t_i}\sqrt{2c(t')}\textrm{d}t'+\phi_i,
\end{equation}
where $\phi_i=\phi(t_i)$. Note that the $\phi(t)$ and $V(\phi(t))$ derived in this way automatically satisfy the background scalar field equation of motion,
\begin{equation}\label{quintessence background EoM}
     \ddot{\phi}+3H\dot{\phi}+\frac{\partial}{\partial\phi}V(\phi)=0,
\end{equation}
as long as $\Lambda(t)$ and $c(t)$ satisfy the conservation equation, \eqref{conservation equation}. Now, because we have assumed that $\phi=\phi(t)$ is monotonic, we can invert it to get $t=t(\phi)$, and consequently $V$ can be reconstructed as
\begin{equation}
    V(\phi)=V(t(\phi)).
\end{equation}
The process is sketched in Figure \ref{fig:fc_quint} (solid line), which allows us to determine the non-linear dynamical evolution by solving
\begin{equation}\label{quintessence EoM}
    \square\phi = \frac{\partial}{\partial\phi}V(\phi).
\end{equation}


\begin{table}[t]
    \centering
    \caption{Parameter sets of four quintessence models.} 
    \label{table:par_quint}
    \begin{tabularx}{\textwidth}{m{50pt}<{\centering}|m{75pt}<{\centering}m{50pt}<{\centering}m{50pt}<{\centering}m{50pt}<{\centering}X<{\centering}}
        \hline
        Model & Parametrization & $w_0$ & $w_a$ & $a_t$ & $\phi(t)$ monotonic? \\
        \hline
        QI & CPL & $-1$ & $0.2$ & -- & yes \\
        QII & CPL & $-0.8$ & $-0.2$ & -- & yes \\
        QIII & Eq.~\eqref{modified CPL} & $-1$ & $0.2$ & $0.7$ & yes\\
        QIV & Eq.~\eqref{modified CPL} & $-1$ & $0.2$ & $0.7$ & no\\
        \hline
    \end{tabularx}
\end{table}

At first glance, the existence of the initial condition $\phi_i$ in Eq.~\eqref{phi integral 1} seems to introduce a degree of arbitrariness into the reconstruction, as different choices of $\phi_i$ may yield different forms of $V(\phi)$. However, this initial condition is inessential for deriving the evolution of the quintessence model. By defining a new scalar field $\varphi=\phi-\phi_i$ and a new potential $U(\varphi)=V(\varphi+\phi_i)=V(\phi)$, it is straightforward to derive that now the action for $\varphi$ is
\begin{equation}
    S=\int d^4x\sqrt{-g}\left[\frac{M^2_*}{2}R-\frac{1}{2}g^{\mu\nu}\partial_\mu\varphi\partial_\nu\varphi-U(\varphi)\right],
\end{equation}
and the EFT parameters relate to this theory by $c(t)=\frac{1}{2}\dot{\varphi}^2$, $\Lambda(t)=U\big(\varphi(t)\big)=U(t)$. As $\varphi_i=\varphi(t_i)=\phi(t_i)-\phi_i=0$, now the integration of $c$ will not involve arbitrary initial condition anymore. 

To ensure that the reconstructed potential $V(\phi)$ is well-defined, attention must be paid to the evolution of the EFT parameters, especially in relation to the behavior of $\phi(t)$. In the simplest and most common case where $\phi(t)$ evolves monotonically, the reconstruction proceeds straightforwardly. This corresponds to $\dot{\phi}(t)\neq0$, and thus $c(t)=\dot{\phi}^2(t)/2>0$, over the entire time interval of interest. As mentioned in Eq.~\eqref{background EFT}, when $\phi$ is minimally coupled, we shall have $2c(t)=\rho_{\textrm{DE}}(t)+P_{\textrm{DE}}(t)=\rho_{\textrm{DE}}(t)(1+w_{\textrm{DE}}(t))$. This leads to $w_{\textrm{DE}}>-1$, as expected for a standard quintessence model. More subtle cases arise when $\dot{\phi}(t)$ is non-negative and vanishes at isolated points. In such scenarios, the potential's gradient $V_\phi = \dot{V} / \dot{\phi}$ would diverge unless $\dot{V}$ also vanishes at the same time. This imposes the condition that the time derivative of $\Lambda(t) = V(t)$ must have zero points coinciding with those of $c(t)=\dot{\phi}^2(t)/2$, i.e., $\dot{\phi}(t_0) = 0 \Rightarrow \dot{\Lambda}(t_0) = 0$, in order for $V_\phi$ to remain finite.

\begin{figure}[t]
\begin{center}
\includegraphics[width=1\textwidth]{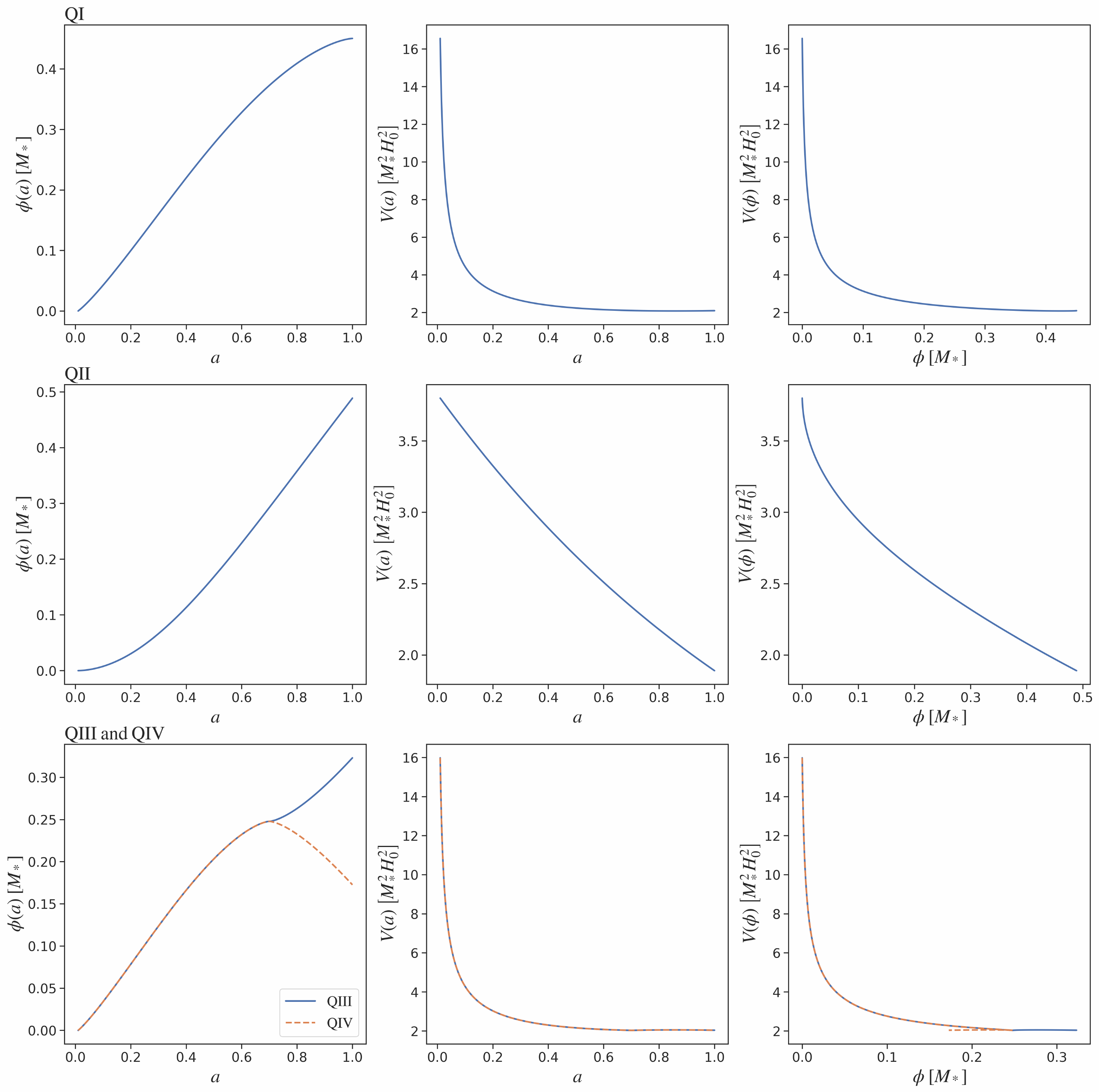}
\caption{A demonstration of the reconstruction process for quintessence models. \textit{The first row}: the reconstruction of model QI. The first two panels show the evolution of $\phi(a)$ and $V(a)$, which are time-dependent quantities determined by the chosen specification of the EFT parameters. By using these results, $V(\phi)$ can be reconstructed (the third panel in this row). Note that $\phi$ and $V$ are in units of $M_*$ and $M_*^2H_0^2$, respectively. \textit{The second row}: the reconstruction of model QII with the same panel configuration as the first row. \textit{The third row}: the reconstruction of model QIII (blue solid line) and QIV (orange dashed line) with the same panel configuration as the first row.}
\label{fig:nu_quint}
\end{center}
\end{figure}

If $\phi(t)$ is assumed to be non-monotonic, additional care is needed. In such cases, the inversion from $\phi(t)$ to $t(\phi)$ may become multi-valued, leading to an unphysical reconstructed potential $V(\phi)$ that is also multi-valued. The only possibility to reconstruct a single-valued smooth potential $V(\phi)$ in this case is that the observed evolutions of $c(t)$ and $\Lambda(t)$ exactly correspond to scenarios in which the scalar field oscillates within a well-defined potential. We conclude that reconstructing a specific DE model, like quintessence, from observations of the EFT parameters does not always work. If the functions $c(t)$ and $\Lambda(t)$ evolve in a way that leads to ill-defined or multi-valued $V(\phi)$, such behavior lies beyond the regime of quintessence models, and the reconstruction method cannot be consistently applied.



As a numerical demonstration, we implement the above reconstruction procedure to 4 specific quintessence models. To achieve this, we first specify the evolution of $c(t)$ and $\Lambda(t)$, or equivalently, parameters related to the evolution of the equation of state, which are presented in Table \ref{table:par_quint}, and the the density parameter at redshift $0$. Here and below, we always fix the mass ratio of matter and DE at redshift $0$ to be $\Omega_m=0.3$ and $\Omega_{\textrm{DE}, 0}=0.7$. We solve for $\phi(a)$ and $V(a)$ in the range where $0.01\leq a\leq1$, with $a$ being the scale factor. The results are displayed in Figure \ref{fig:nu_quint}. Models QI and QII correspond to the class of freezing dark energy and thawing dark energy, respectively. For these two models, the equation of state of dark energy follows the CPL parametrization
\begin{equation}
    w_{\textrm{DE}}(a)=w_0+w_a(1-a).
\end{equation}
Models QIII and QIV are displayed to illustrate the multi-value problem mentioned in the last paragraph. They follow a slightly modified version of the CPL parametrization, which can be specified as
\begin{equation}\label{modified CPL}
    w_{\textrm{DE}}(a)=w_0+w_a|a_t-a|,
\end{equation}
with $a_t$ being a constant. The difference between these two models is that for model QIV, the scalar $\phi(t)$ will move backward after $a=a_t=0.7$. From the dashed orange line in the lower right panel of Figure \ref{fig:nu_quint}, the parameter sets of this model do lead to reconstructing a multi-valued $V(\phi)$, consistent with the expected behavior for non-monotonic $\phi(t)$.

The above method can be straightforwardly extended to scalar-tensor theories. This class of theories involves non-zero energy transitions between dark energy and matter, which is well-suited to explain recent observations that the dark energy equation of state has undergone a phantom-crossing event \cite{STT_phantom_crossing1, STT_phantom_crossing2}. These theories can be specified as
\begin{equation}
    K=-\frac{1}{2}X-V(\phi),\quad G_3=0,\quad G_4=\xi(\phi).
\end{equation}
In this case, the only difference is the inclusion of an additional EFT parameter $f(t)=\xi(\phi(t))=\xi(t)$, which captures the non-minimal coupling characteristic of such theories. The following procedure remains essentially the same: by fixing $f(t)$, $\Lambda(t)$, and $c(t)$, we will derive $\phi(t)$, $V(t)$, and $\xi(t)$, and thereby $V(\phi)$ and $\xi(\phi)$. The reconstruction processes of scalar-tensor theory are also sketched in Figure \ref{fig:fc_quint} (solid line + dashed line). The numerical implementations for these models are omitted as they would straightforwardly replicate the quintessence reconstruction (apart from an additional $\xi(\phi)$-dependent term).


\section{Reconstruction of \boldmath \texorpdfstring{$k$}{}-essence}
\label{sec:kessence}

In this section, we pay our attention to the $k$-essence model. For this model, we have
\begin{equation}\label{kessence}
    K=K(\phi, X),\quad G_3=0,\quad G_4=1.
\end{equation}
Previous papers on reconstructing this class of models usually made some strong assumptions about the form of $K(\phi, X)$. For instance, Refs.~\cite{kessence_reconstruct1, kessence_reconstruct4} mainly focused on the reconstruction for purely kinetic $k$-essence, or  $K=K(X)$, \cite{kessence_reconstruct3} reconstruct $K=K(q(\phi)X)$ by additionally fixing $\phi(t)\propto t$. In Ref.~\cite{kessence_reconstruct2} the authors studied models such as $K=f(X)-V(\phi)$ or $V(\phi)f(X)$. After specifying the $X$-dependence in these models, they reconstructed the $\phi$-dependence using a method similar to the methods in the papers listed at the beginning of the last section.\footnote{Note that, like there, here we are not attempting to provide an exhaustive list of references for reconstruction works on the $k$-essence model.} Going beyond these approaches, in the following part of this section, we will show that our reconstruction framework can be implemented for more complicated $k$-essence theories.

We first derive the general formulae for $k$-essence. Substituting the definition of Eq.~\eqref{kessence} into Eq.~(\ref{EFT relation}), we get
\begin{equation}\label{kessence relation}
\begin{aligned}
    \Lambda(t)&=-K_X(\phi, X)\dot{\phi}^2-K(\phi, X)=K_X(\phi, X)X-K(\phi, X),\\
    c(t)&=-K_X(\phi, X)\dot{\phi}^2=K_X(\phi, X)X,\\
    M_2^4(t)&=K_{XX}(\phi, X)\dot{\phi}^4=K_{XX}(\phi, X)X^2.
\end{aligned}
\end{equation}
For this class of models, we can always redefine the scalar field as $\phi=\phi(\varphi)$ to transform the function $K$ into a new form,
\begin{equation}
    K(\phi,X_\phi)\rightarrow \tilde{K}(\varphi,X_\varphi)=K\left(\phi(\varphi), X_\varphi\left[\frac{\textrm{d}}{\textrm{d}\varphi}\phi(\varphi)\right]^2\right),
\end{equation}
in which $X$ is defined in the same way as above, but with a subscript $_\phi$ or $_\varphi$ to label which scalar field is used. In this process, the EFT parameters $\Lambda(t)$, $c(t)$, $M_2^4(t)$ do not change. This field redefinition freedom means that we can fix $\phi(t)$: out of the infinite number of $\phi(t)$ functions that give the same background evolution, we can choose a convenient one. With $\phi(t)$ fixed, so is $X(t)$, and then Eq.~\eqref{kessence relation} can be used to determine $K(t)$, and $K_X(t)$, $K_{XX}(t)$, \textcolor{black}{as well as $K_\phi(t)$, from the relation $\dot{K}=K_\phi\dot{\phi}+K_X\dot{X}$}.\footnote{A Similar operation can also be done for the quintessence model. However, via field redefinition, the Lagrangian is transformed into $L=\omega(\varphi)X_\varphi-V(\varphi)$ with $\omega(\varphi)=\left[\textrm{d}\phi(\varphi)/\textrm{d}\varphi\right]^2$ at this time. Now we still have two functions, namely $\omega$ and $V$, to be reconstructed, which means field redefinition does not notably simplify the reconstruction process in this model compared to the last section.} We note that this does not fully reconstruct $K(\phi, X)$. This is because to do so, we need to know the values of $K$ in the entire $\phi$ -- $X$ plane (or at least a 2D region therein), while the above process only gives the values of $K$ on a parameterized 1D curve $(\phi(t), X(t))$ in this plane.\footnote{\textcolor{black}{But as we will briefly mentioned in the last section, for cosmological applications, we can retrieve "just enough" information from this 1D curve.}} To fully reconstruct $K(\phi, X)$, we need to make some additional assumptions.

\begin{figure}[t]
\begin{center}
\includegraphics[width=1\textwidth]{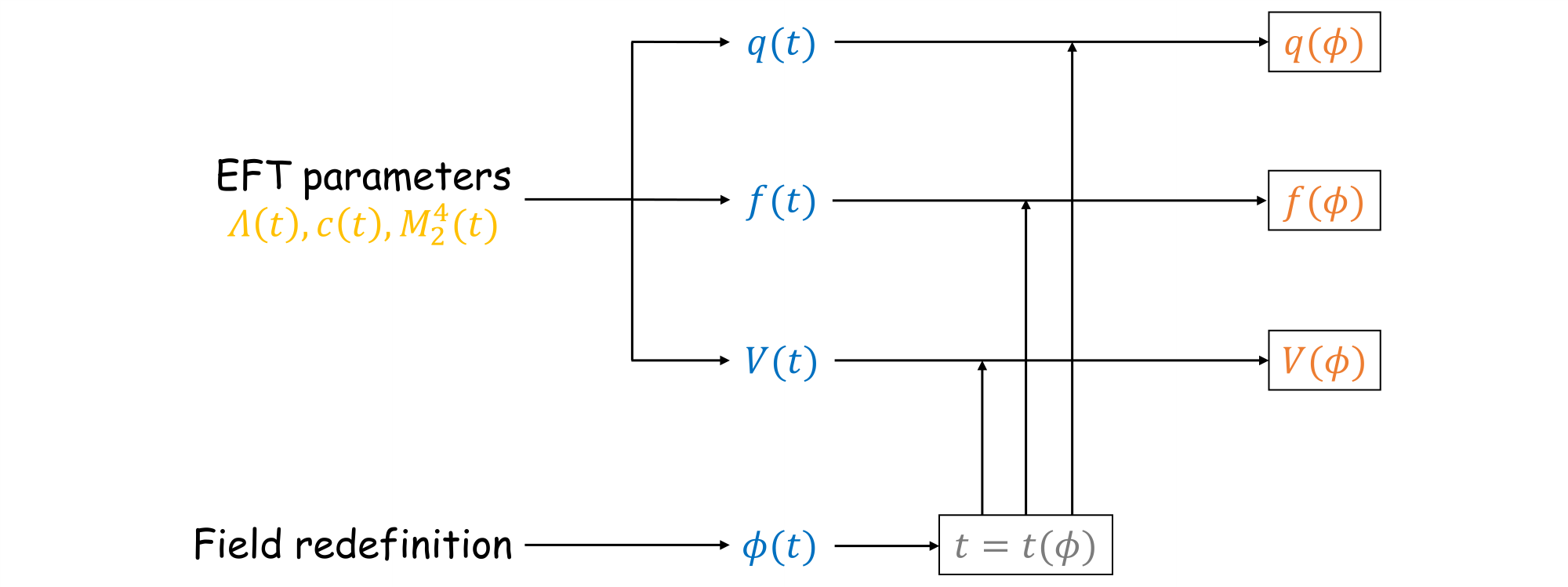}
\caption{A sketch of the reconstruction scheme for the $k$-essence models with $K(\phi, X)=V(\phi)+f(\phi)X+q(\phi)X^2$. The colors of the quantities are the same as Figure \ref{fig:fc_quint}.}
\label{fig:fc_kessence1}
\end{center}
\end{figure}

An example of simplifying assumptions is $K(\phi, X)=f(\phi)X+q(\phi)X^2$ (which was first proposed in Eq.~(2.3) of Ref.~\cite{kessence_theory1}), with $f$ and $q$ being unknown functions of $\phi$ only, for which
\begin{eqnarray}
    \Lambda(t) &=& q(\phi)X^2,\nonumber\\
    c(t) &=& f(\phi)X + 2q(\phi)X^2,\\
    M_2^4(t) &=& 2q(\phi)X^2.\nonumber
\end{eqnarray}
We note that this is a special case where $2\Lambda(t)= M_2^4(t)$, so we \textit{cannot} parameterize $\Lambda(t)$ and $M_2^4(t)$ differently. Hence, only $\Lambda$ and $c$ are independent. If we fix the form of $\phi$, say $\phi(t)\propto t$, and thereby $X(t)\propto-1$, we shall fully determine $f(t)$ and $q(t)$, and thus $f(\phi)$ and $q(\phi)$.

We can further assume $K(\phi, X)=V(\phi)+f(\phi)X+q(\phi)X^2$, by adding a potential $V(\phi)$ to the model. For this model, we have
\begin{eqnarray}\label{kessence relation 0}
    c(t)-\Lambda(t) &=& V(\phi)+f(\phi)X+q(\phi)X^2,\nonumber\\
    c(t) &=& f(\phi)X + 2q(\phi)X^2,\\
    M_2^4(t) &=& 2q(\phi)X^2,\nonumber
\end{eqnarray}
in which case all three equations are needed to reconstruct $V(\phi)$, $f(\phi)$, and $q(\phi)$. The reconstruction process is sketched as Figure \ref{fig:fc_kessence1}. Note that now we \textit{cannot} include one more term such as $p(\phi)X^3$ into $K(\phi, X)$, as in that case we would have 4 unknown functions to solve from 3 independent equations, which leads to under-constraints. 

Before proceeding, we digress briefly to remark that the reconstruction framework proposed in Refs.~\cite{EFT_reconstruct1, EFT_reconstruct2, EFT_reconstruct3, EFT_reconstruct4} is similar to what we have described above, except that the theories they reconstructed include more terms: they need to first fix $\phi(t)$ to be proportional to $t$ through a field redefinition, and then subscribe the definition of $\phi(t)$ into the EFT parameters in the action. The theories reconstructed this way can be considered as the first few terms of the Taylor expansion with respect to $X$ of a full theory, whose expansion parameters are functions of $\phi$. These papers also mentioned that they can reconstruct higher-order expansion terms by utilizing higher-order EFT parameters, which resolves the under-constraining problem seen in the last paragraph. In summary, their reconstruction framework mainly focuses on reconstructing these $\phi$-dependent functions, and they can better describe the general DE theory when expanding more terms. As a comparison, the reconstruction method proposed in this paper usually requires a prior narrowing-down of the general DE theory, Eq.~\eqref{GG action}. However, except for theories with the form of $K(\phi, X)=V(\phi)+f(\phi)X+q(\phi)X^2$ we are discussing here, the reconstructions of all the other DE models are mainly based on solving the differential equation~\eqref{EFT relation} to obtain the evolution of $\phi(t)$ and $X(t)$ (as we will see in the following part of this paper), and thus we can derive both $\phi$ and $X$-dependence of DE theories.

Now return to our discussion of the reconstruction framework. As shown in~\cite{k_essence_stability}, in most cases, the phenomenon of phantom crossing, i.e., the equation of state $w_{\textrm{DE}}$ crossing $-1$ at some point $t_\textrm{p}$, poses theoretical obstacles for minimally coupled scalar-field dark energy scenarios like $k$-essence. Therefore, here we analyze the behavior of the reconstruction results through the phantom crossing. \textcolor{black}{In Ref.~\cite{k_essence_stability}, the author first assumed a positive sound speed for DE, i.e., the classical gradient stability, and then analyzed the evolution of the background equation of motion of the scalar field. They found that} the equation of state $w_{\textrm{DE}}$ of a $k$-essence theory $K(\phi, X)$ can stably cross $-1$ only when at this time it satisfies
\begin{equation}
\begin{aligned}
    \rho_{\textrm{DE},\phi}=0,\quad
    \rho_{\textrm{DE},X}=0,\quad
    P_{\textrm{DE},X}&=0\\
    \dot{\phi}\left[\left(\rho^2_{\textrm{DE},X\phi}-\rho_{\textrm{DE},XX}\rho_{\textrm{DE},\phi\phi}\right)\dot{\phi}-3H\rho_{\textrm{DE},XX}P_{\textrm{DE},\phi}\right]&>0.
\end{aligned}
\end{equation}
More details can be found in Section III of Ref.~\cite{k_essence_stability}. From the relation $2c(t)=\rho_{\textrm{DE}}(t)+P_{\textrm{DE}}(t)=\rho_{\textrm{DE}}(t)\left[1+w_{\textrm{DE}}(t)\right]$, the function $c(t)$ will change its sign at the phantom crossing, and so $c(t_\textrm{p})=0$. The square of the speed of sound for $k$-essence can be written as
\begin{equation}
    c_s^2=\frac{c}{c+2M_2^4}.
\end{equation}
As a result, to guarantee that $c_s^2$ stays non-negative and no greater than $1$ to ensure the stability of the dark energy, the function $M_2^4(t)$ should have the same sign as $c(t)$, and $M^4_2(t_\textrm{p})=0$\footnote{This is a loose requirement. In practice, we also need to place constraints on how fast $M_2^4$ approaches $0$ as $c\rightarrow0$.}. From Eq.~\eqref{kessence relation 0}, this leads to $f(t_\textrm{p})=0$ and $q(t_\textrm{p})=0$ as we have assumed that $\dot{\phi}$ does not cross $0$. By using the fact that $\rho_{\textrm{DE}}=c+\Lambda$, $P_{\textrm{DE}}=c-\Lambda$, and the relation~\eqref{kessence relation 0}, we can derive that at time $t_\textrm{p}$, $P_{\textrm{DE},X}=0$ and $\rho_{\textrm{DE},X}=0$. Next, from 
\begin{equation}
    \frac{\textrm{d}}{\textrm{d}t}\rho_{\textrm{DE}}(\phi, X)=\rho_{\textrm{DE},\phi}\dot{\phi} + \rho_{\textrm{DE},X}\dot{X},
\end{equation}
$\rho_{\textrm{DE},\phi}$ will vanish at $t_\textrm{p}$ as long as $\dot{\phi}$ does not vanish at that time, because $\dot{\rho}_{\rm DE}=0$ then. In addition, we have
\begin{equation}
\begin{aligned}
    &\dot{\phi}\left[\left(\rho^2_{\textrm{DE},X\phi}-\rho_{\textrm{DE},XX}\rho_{\textrm{DE},\phi\phi}\right)\dot{\phi}-3H\rho_{\textrm{DE},XX}P_{\textrm{DE},\phi}\right]=\rho^2_{\textrm{DE},X\phi}\dot{\phi}^2\\
    =&\left(f_\phi-6q_\phi\dot{\phi}^2\right)^2\dot{\phi}^2=\left(\dot{f}-6\dot{q}\dot{\phi}^2\right)^2.
\end{aligned}
\end{equation}
For the general case where $\dot{f}\neq6\dot{q}\dot{\phi}^2$, the above formula will be positive at $t_\textrm{p}$. This analysis demonstrates that the reconstructed $K$ in this way automatically satisfies the \textcolor{black}{classical} stability condition mentioned in \cite{k_essence_stability}. As a result, this reconstruction formalism provides a straightforward way to obtain a concrete $k$-essence Lagrangian stable at the phantom-crossing.\footnote{We examine the stability of the $k$-essence here to make the reconstruction of a model with phantom-crossing in the next paragraph seem more reasonable. We note here that there are also some papers \cite{k_essence_stability2, k_essence_stability3, k_essence_stability4} which claimed that all $k$-essence models are unstable at the phantom crossing by analyzing the linear fluid equations. \textcolor{black}{These papers also start from the assumption of classical gradient stability, but different from \cite{k_essence_stability}, they also examined the evolution of the spatial perturbations of the physical quantities, like $\delta\phi(x^\mu)$ and $\delta\rho(x^\mu)$. They found that even when the gradient stability is satisfied, the spatial perturbations will still diverge at the phantom crossing event.} However, this conclusion does not affect the reconstruction process, so we can still implement our framework on a phantom-crossing $k$-essence model.}


\begin{table}[t]
    \centering
    \caption{Parameter sets of three $k$-essence models with $K(\phi, X)=V(\phi)+f(\phi)X+q(\phi)X^2$.} 
    \label{table:par_k1}
    \begin{tabularx}{\textwidth}{X<{\centering}|X<{\centering}X<{\centering}X<{\centering}}
        \hline
        Model & Parametrization & $w_0$ & $w_a$ \\
        \hline
        KI-I & CPL & $-1.0$ & $-0.2$ \\
        KI-II & CPL & $-0.8$ & $-0.2$ \\
        KI-III & CPL & $-0.8$ & $-0.4$ \\
        \hline
    \end{tabularx}
\end{table}

\begin{figure}[t]
\begin{center}
\includegraphics[width=1\textwidth]{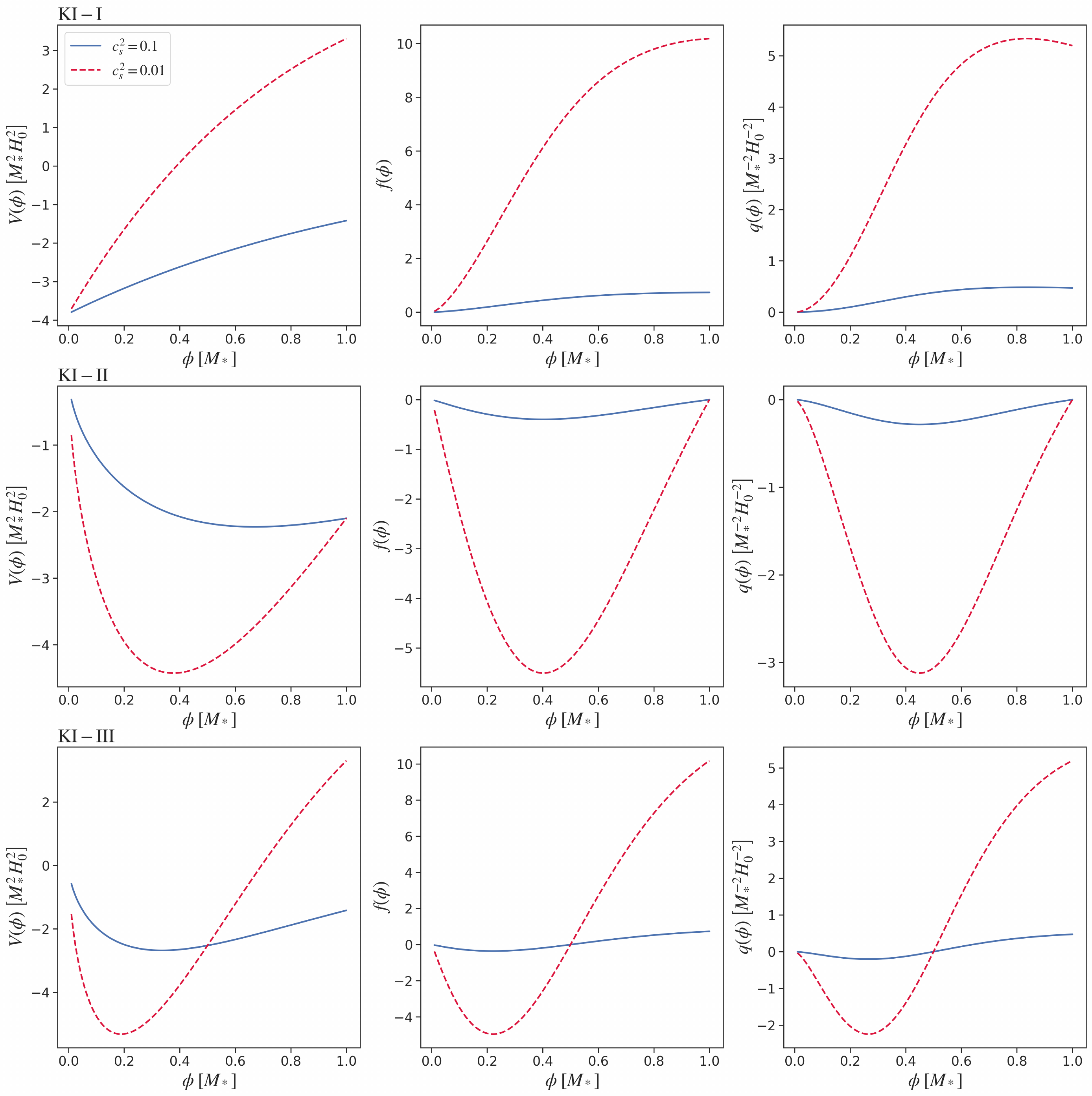}
\caption{Numerical results of reconstructing $k$-essence models with $K(\phi, X)=V(\phi)+f(\phi)X+q(\phi)X^2$. \textit{The first row}: The reconstruction of model KI-I. From left to right, we exhibit the functional forms of $V(\phi)$, $f(\phi)$, and $q(\phi)$, in units of $M_*^2H_0^2$, $1$, and $M_*^{-2}H_0^{-2}$, respectively. As we fix $\phi(a)= aM_*$, the curves in these panels equivalently correspond to the evolution of $V(a)$, $f(a)$, and $q(a)$, which are time-dependent quantities determined by given evolutions of the EFT parameters. Each panel presents the results with $c_s^2=0.1$ (blue solid line) and $c_s^2=0.01$ (red dashed line). \textit{The second row} and \textit{the third row} show the reconstruction of model KI-II and KI-III with the same panel configuration as the first row.} 
\label{fig:nu_kessence1}
\end{center}
\end{figure}

Now we give some numerical examples of $k$-essence reconstruction, as we did for quintessence. The parameter sets are summarized in Table \ref{table:par_k1}. Models KI-I and KI-II are DE models without phantom crossing, whose equation of state $w_{\textrm{DE}}$ is less than or greater than $-1$ all the time. On the other hand, model KI-III provides a concrete example of a stable $k$-essence model with a phantom-crossing event. For all these models, we fix $\phi(t)$ to be proportional to the scale factor, $\phi(t)=a(t) M_*$, where we have kept the dimension of the scalar field. The results are displayed in Figure \ref{fig:nu_kessence1}. Note that for these models, we also specify the evolution of $c(a)$ and $\Lambda(a)$ by giving the value of $w_0$ and $w_a$ (and set $\Omega_m=0.3$ and $\Omega_{\textrm{DE}, 0}=0.7$), and specify the evolution of $M_2^4(a)$ by fixing the value of the sound speed $c_s^2$. From Eq.~\eqref{kessence relation 0}, the constant sound speed will lead to $fX$ and $qX^2$ being proportional to each other. As we have fix $\phi(t)=a(t) M_*$, this means $f$ is proportional to $aHq$. This explains the similarity between the second and the third panel in each row.

A more interesting and general simplifying assumption is that $K(\phi, X)$ is factorizable, i.e., $K(\phi, X)=q(\phi)h(X)$, with $q$ and $h$ being unknown functions. We can still take the field redefinition $\phi=\phi(\varphi)$. However, in this process, the function $K$ will transform as:
\begin{equation}
    K(\phi, X_\phi)\rightarrow q(\phi(\varphi))h\left(X_\varphi\left[\frac{\textrm{d}}{\textrm{d}\varphi}\phi(\varphi)\right]^2\right).
\end{equation}
This final expression no longer allows for the separation of variables, which makes the situation more complicated. Therefore, instead of specifying the evolution of $\phi(t)$, we shall derive it by solving the differential equations. 

\begin{figure}[t]
\begin{center}
\includegraphics[width=1\textwidth]{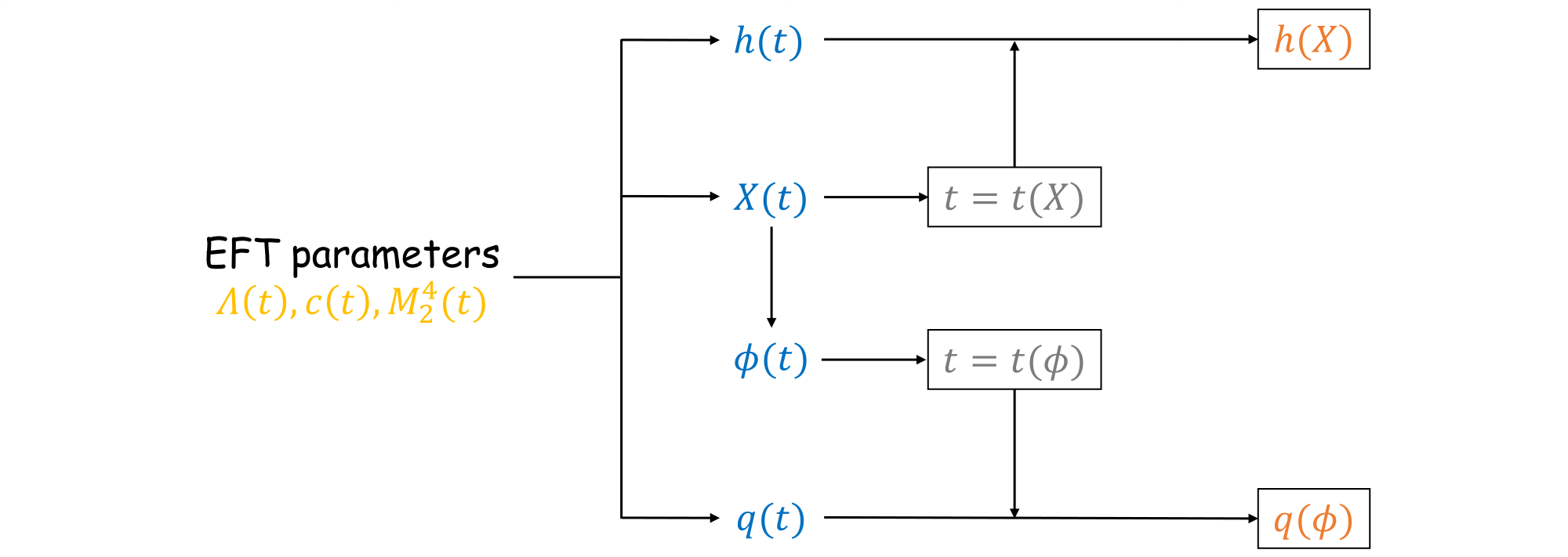}
\caption{A sketch of the reconstruction scheme for the $k$-essence models with $K(\phi, X)=q(\phi)h(X)$. The colors of the quantities are the same as Figure \ref{fig:fc_quint}.}
\label{fig:fc_kessence2}
\end{center}
\end{figure}

From the relation $\dot{h}\left(X(t)\right)=h_X\dot{X}$, we have
\begin{equation}
    h_X=\frac{\dot{h}}{\dot{X}},\quad h_{XX}=\frac{1}{\dot{X}^2}\left(\ddot{h}-\dot{h}\frac{\ddot{X}}{\dot{X}}\right).
\end{equation}
Now, Eq.~(\ref{kessence relation}) can be rewritten as
\begin{equation}\label{kessence relation 1}
\begin{aligned}
    c(t)-\Lambda(t)&=qh,\\ c(t)&=q\dot{h}\frac{X}{\dot{X}},\\
    M_2^4(t)&=q\frac{X^2}{\dot{X}^2}\left(\ddot{h}-\dot{h}\frac{\ddot{X}}{\dot{X}}\right).
\end{aligned}
\end{equation}
We can solve $q$ from the first equation, and substitute it into the other two equations. Then, by taking the time derivative of the second equation and substituting it into the third equation, the second-order derivatives of $h$ and $X$ cancel out. The final result is
\begin{equation}\label{kessence relation 2}
\begin{aligned}
    q&=\frac{c-\Lambda}{h},\\
    \frac{\dot{h}}{h}&=\frac{c}{c-\Lambda}\frac{\dot{X}}{X},\\
    \frac{\dot{X}}{X}&=\left(c-\Lambda\right)\frac{\textrm{d}}{\textrm{d}t}\left[\frac{c}{c-\Lambda}\right]\bigg/\left(M_2^4-\frac{c\Lambda}{c-\Lambda}\right).
\end{aligned}
\end{equation}
From the last equation, we get
\begin{equation}\label{k X integral}
    X(t)=X_i\exp\left[\int^t_{t_i}\left(\left(c-\Lambda\right)\frac{\textrm{d}}{\textrm{d}t'}\left[\frac{c}{c-\Lambda}\right]\bigg/\left(M_2^4-\frac{c\Lambda}{c-\Lambda}\right)\right)\textrm{d}t'\right],
\end{equation}
where $X_i=X(t_i)$. Thus, if $X_i\neq0$, then $X$ will never equal zero, which means $\dot{\phi}$ will not change sign. Subsequently, we can compute $\phi(t)$ by
\begin{equation}
    \phi(t)=\int^t_{t_i}\sqrt{-X(t')}\textrm{d}t'+\phi_i.
\end{equation}
Similarly, from the second equation of Eq.~\eqref{kessence relation 2}, the evolution of $h$ should be
\begin{equation}\label{k h integral}
    h(t)=h_i\exp\left[\int^t_{t_i}\left(c\frac{\textrm{d}}{\textrm{d}t'}\left[\frac{c}{c-\Lambda}\right]\bigg/\left(M_2^4-\frac{c\Lambda}{c-\Lambda}\right)\right)\textrm{d}t'\right].
\end{equation}
Then from the first equation of Eq.~\eqref{kessence relation 2}, we can derive the evolution of $q(t)$. The remaining part is similar to the quintessence case above. We know that $\phi(t)$ is monotonic in $t$, and by further assuming that $X(t)$ is also monotonic, we can solve for $t=t(\phi)$ and $t=t(X)$ and substitute them into $q(t)$ and $h(t)$, respectively, to obtain $q(\phi)$ and $h(X)$ and thus $K(\phi, X)$. The reconstruction process is sketched in Figure \ref{fig:fc_kessence2}.

\begin{table}[t]
    \centering
    \caption{Parameter sets of three $k$-essence models with $K(\phi, X)=q(\phi)h(X)$.} 
    \label{table:par_k2}
    \begin{tabularx}{\textwidth}{X<{\centering}|X<{\centering}X<{\centering}X<{\centering}}
        \hline
        Model & Parametrization & $w_0$ & $w_a$ \\
        \hline
        KII-I & CPL & $-0.8$ & $-0.2$ \\
        KII-II & CPL & $-1.0$ & $-0.2$ \\
        \hline
    \end{tabularx}
\end{table}

Now we need to discuss the influence of the initial condition, i.e., $\phi_i=\phi(t_i)$, $X_i=X(t_i)$, and $h_i=h(t_i)$. First, similar to quintessence, $\phi_i$ is insignificant, as we can always remove its effects by defining a new scalar field $\varphi=\phi-\phi_i$. Second, the initial condition of $X(t)$ does not matter either. This is because by using $X_i$, we can always do a field rescaling as
\begin{equation}
    \phi\rightarrow\tilde{\phi}=\frac{\phi}{\sqrt{-X_i/M_*^2H_0^2}},\quad X=-\dot{\phi}^2 \rightarrow\tilde{X}=-\frac{X}{X_i/M_*^2H_0^2},
\end{equation}
and the initial condition for $\tilde{X}$ will just be $\tilde{X}_i=-M_*^2H_0^2$ from the above definition. Just like before, the factor $M_*^2H_0^2$ encodes the dimension of $X$. With these, the Lagrangian of $k$-essence transforms to $\tilde{K}(\tilde{\phi},\tilde{X})=K(\sqrt{-X_i/M_*^2H_0^2}\tilde{\phi}, -X_i\tilde{X}/M_*^2H_0^2)$ whose corresponding evolution, both linear and non-linear, remains identical to the original $K(\phi, X)$. It is straightforward to verify that Eq.~\eqref{kessence relation 0} now can be written as
\begin{equation}
\begin{aligned}
    \Lambda(t)&=\tilde{K}_{\tilde{X}}(\tilde{\phi},\tilde{X})\tilde{X}-\tilde{K}(\tilde{\phi},\tilde{X}),\\
    c(t)&=\tilde{K}_{\tilde{X}}(\tilde{\phi},\tilde{X})\tilde{X},\\
    M_2^4(t)&=\tilde{K}_{\tilde{X}\tilde{X}}(\tilde{\phi},\tilde{X})\tilde{X}^2.
\end{aligned}
\end{equation}
Then, without incurring additional complications, we can entirely replicate the previous procedure to reconstruct $\tilde{X}$. This implies that we can always specify $X_i=-M_*^2H_0^2$, without loss of generality. Finally, from Eq.~\eqref{k h integral} and the first line of Eq.~\eqref{kessence relation 2}, we notice that the evolution of $h(t)/h_i$ and $h_iq(t)$ is independent of $h_i$, once we have specified the EFT parameters. As a result, the forms of $h(X)/h_i$ and $h_iq(\phi)$ are also independent of $h_i$. Therefore, the initial condition of $h(t)$ is irrelevant for determining $K(\phi, X)=q(\phi)h(X)=\left[h_iq(\phi)\right]\cdot\left[h(X)/h_i\right]$. We conclude that, for the factorizable $k$-essence, the reconstruction procedure is fully independent of the initial conditions. 

\begin{figure}[p]
\begin{center}
\includegraphics[width=1\textwidth]{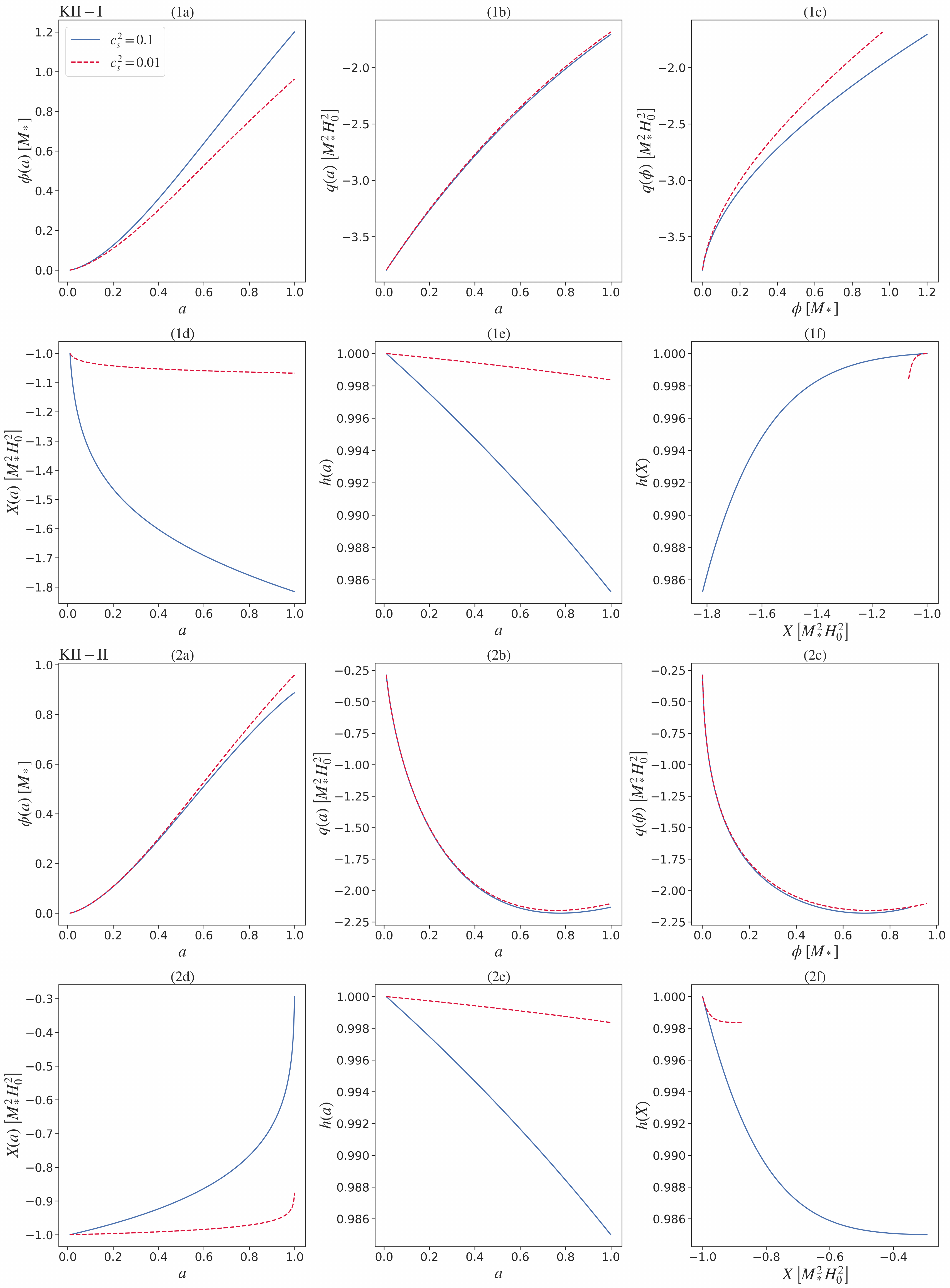} 
\caption{An example of reconstructing $k$-essence models with $K(\phi, X)=q(\phi)h(X)$. Panels (1a) and (1b) show the evolution of $\phi(a)$ and $q(a)$ for model KII-I, which are time-dependent quantities determined by given specifications of the EFT parameters. Then we invert $\phi(a)$ and reconstruct $q(\phi)$ (panel 1c). Similarly, panels (1d) and (1e) show the evolution of $X(a)$ and $h(a)$, and the reconstruction of $h(X)$ is shown in panel (1f). We set $h$ to be dimensionless, while $q$, $\phi$, $X$ are in units of $M_*^2H^2_0$, $M_*$, and $M_*^2H^2_0$, respectively. Panels (2a)--(2f) are the same as panels (1a)--(1f), except that they exhibit the reconstruction results for model KII-II.} 
\label{fig:nu_kessence2}
\end{center}
\end{figure}
\enlargethispage{\baselineskip}

Unlike the previous example, the factorizable $k$-essence model here becomes incorrect if we demand it to keep a positive sound speed during the phantom crossing. We can see this from Eq.~\eqref{kessence relation 2}. First, assume \textcolor{black}{the classical gradient stability, i.e.,} the sound speed squared satisfies $0<c^2_s\leq1$, and that, without loss of generality, the dark energy equation of state monotonically increases around the phantom crossing. As before, $c(t)$ and $M^4_2(t)$ should both change its sign at the phantom crossing, and $c(t_\textrm{p})=0$ and $M_2^4(t_\textrm{p})=0$ with $t_\textrm{p}$ being the time when $w_\textrm{DE}$ crosses $-1$. At this time, the denominator in the integral~\eqref{k X integral} vanishes, possibly leading to the integral diverging. But even if Eq.~\eqref{k X integral} remains finite, the phantom crossing will also lead to other problems. By using Eq.~\eqref{background EFT}, we can derive that 
\begin{equation}
\begin{aligned}
    &c-\Lambda=P_{\textrm{DE}}=w_{\textrm{DE}}\rho_{\textrm{DE}}<0,\\
    &\frac{\textrm{d}}{\textrm{d}t}\left[\frac{c}{c-\Lambda}\right]=\frac{1}{2}\frac{\textrm{d}}{\textrm{d}t}\left[\frac{1+w_{\textrm{DE}}}{w_{\textrm{DE}}}\right]=\frac{1}{2}\frac{\textrm{d}}{\textrm{d}t}\left[\frac{1}{w_{\textrm{DE}}}+1\right]<0,\\
    &-\frac{\Lambda}{c-\Lambda}=-\frac{1-w_{\textrm{DE}}}{2w_{\textrm{DE}}}>0.
\end{aligned}
\end{equation}
From the last equation of Eq.~\eqref{kessence relation 2}, this implies that $\dot{X}$ will change sign before and after the phantom crossing event. On the other hand,
\begin{equation}
    \frac{c}{c-\Lambda}=\frac{1+w_{\textrm{DE}}}{2w_{\textrm{DE}}},
\end{equation}
which will also change sign when $w_{\textrm{DE}}$ crosses $-1$. According to the second equation of Eq.~\eqref{kessence relation 2}, this indicates that during the phantom crossing, $\dot{h}$ will preserve its sign, and thus $h(t)$ is monotonic. In such cases, the inversion from $X(t)$ to $t(X)$ yields a multi-valued solution, resulting in a multi-valued reconstructed function $h(X)$. Consequently, the reconstruction frame~\eqref{kessence relation 2} becomes invalid, as long as we ask for a positive sound speed. These results align with the stability analysis presented in Section III.C of~\cite{factorizable_k_essence_stability}.

The parameter sets of the toy models we intend to reconstruct as examples of the factorizable $k$-essence are exhibited in Table \ref{table:par_k2}, and here we again choose $\Omega_m=0.3$ and $\Omega_{\textrm{DE}, 0}=0.7$ and specify the evolution of $M_2^4(a)$ by fixing the value of the sound speed $c_s^2$. Because of the previous analysis, we will only consider models without phantom crossing in this case. The final results are displayed in Figure \ref{fig:nu_kessence2}. For model KII-II, when $a=1$, the EoS $w_{\textrm{DE}}=-1$. This means that for this model, $c(a=1)=0$. As we have assumed the constant sound speed, the $M_2^4$ parameter should also vanish at this time. As a result, the denominator of the third equation of Eq.~\eqref{kessence relation 2} equals zero at $a=1$, while at the same time the numerator does not vanish. This explains the drastic rise of $X(a)$ near $a=1$ in the model KII-II. 

In addition, we note that in the panels 1f and 2f of Figure \ref{fig:nu_kessence2}, the graphic of the reconstructed $h(X)$ with $c_s^2=0.01$ only occupies a narrower region in coordinate space, compared to the model with $c_s^2=0.1$. This can be attributed to the fact that for a $k$-essence model with a smaller sound speed, its corresponding $M^4_2(t)$ parameter will be larger, leading to a smaller $\dot{X}$ and $\dot{h}$ from the second and the last equation of Eq.~\eqref{kessence relation 2}. As a result, in a given range of the scale factor, the evolution scale of the $X(a)$ and $h(a)$ will be smaller, further making the reconstructed $h(X)$ span a reduced domain. This unveils a feature of our reconstruction framework: the methodology is inherently limited to reconstructing one continuum part of the DE Lagrangian in this framework. When the observed EFT parameters lead to a smaller dynamical range of $X(t)$ or $\phi(t)$ or other functions, the reconstructible domain of the DE Lagrangian contracts. This, however, should not be an issue in practice, because this reduced parameter space would still allow us to study the structure formation.

\section{Reconstruction of generalized cubic Galileon with shift symmetry}
\label{sec:galileon}

In this section, we shall focus on a subclass of generalized cubic Galileon models whose Lagrangian is independent of $\phi$, which means
\begin{equation}\label{cG}
    K=K(X),\quad G_3=G_3(X),\quad G_4=1.
\end{equation}
Several broadly-studied models can be obtained by specifying the functional form of $K(X)$ and $G_3(X)$, like the covariant cubic Galileon (with linear $K$ and $G_3$, e.g., the $L_2$ and $L_3$ mentioned in \cite{cubic_Galileon1, cubic_Galileon2, tracker_linear}), GCCG (with power-law $K$ and $G_3$, e.g., \cite{tracker_power, GCCG1, GCCG2}), and Galileon ghost condensation (with $K\sim X+X^2$ and a linear $G_3$, e.g., \cite{GGC1, GGC2}). There are also efforts devoted to the reconstruction of this class of models. In \cite{cG_reconstruct3, cG_reconstruct4, cG_reconstruct5, cG_reconstruct6}, the authors proposed a method to get a designer Horndeski model, by demanding the background evolution to be identical to a $\Lambda$CDM cosmology and specifying the relation between $X$ and the evolution of the Hubble parameter. Besides, \cite{cG_reconstruct7} fixed the exact form of $K(X)$ and then used the expansion history to reconstruct only the $G_3(X)$ (where they needed to exploit the property of the tracker solution, which we will discuss below). Unlike these works, in this section, we will show that with our reconstruction method, this theory can actually be reconstructed \textit{without} assuming a $\Lambda$CDM background or a specific functional form of $K(X)$ or $G_3(X)$.

\begin{figure}[t]
\begin{center}
\includegraphics[width=1\textwidth]{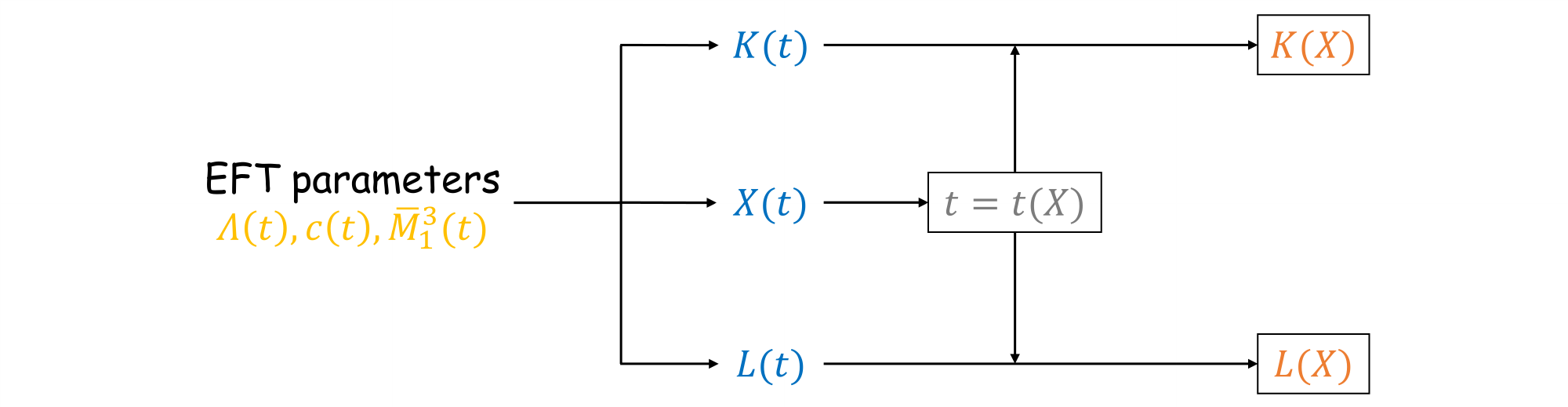}
\caption{A sketch of the reconstruction scheme for the generalized cubic Galileon model with shift symmetry. The colors of the quantities are the same as Figure \ref{fig:fc_quint}.}
\label{fig:fc_cG1}
\end{center}
\end{figure}

For models specified in Eq.~\eqref{cG}, Eq.~\eqref{EFT relation} can be rewritten as
\begin{equation}\label{cG relation 0}
\begin{aligned}
    \Lambda(t)&=\dot{\phi}^2G_{3X}(3H\dot{\phi}+\ddot{\phi})-K_X\dot{\phi}^2-K,\\
    c(t)&=\dot{\phi}^2G_{3X}(3H\dot{\phi}-\ddot{\phi})-K_X\dot{\phi}^2,\\
    M^4_2(t)&=\frac{1}{2}\dot{\phi}^2G_{3X}(3H\dot{\phi}+\ddot{\phi})-3H\dot{\phi}^5G_{3XX}+K_{XX}\dot{\phi}^4,\\
    \bar{M}^3_1(t)&=-2\dot{\phi}^3G_{3X}.
\end{aligned}
\end{equation}
To implement our reconstruction framework, it is convenient to define an auxiliary function
\begin{equation}
    L(X)=H_0(-X)^{1/2}\frac{G_{3X}(X)}{K_X(X)},
\end{equation}
where the Hubble constant, $H_0$, is multiplied to ensure that this function is dimensionless. As in the previous sections, to reconstruct $K(X)$ and $L(X)$, we need to derive the evolutions of $K(t)$, $L(t)$, and $X(t)$. Note that there are four equations in Eq.~\eqref{cG relation 0}, which is more than the number of unknown functions. Therefore, in the rest of this section, we always assume that only $c(t)$, $\Lambda(t)$ and $\bar{M}_1^3(t)$ evolve independently, while the evolution of $M_2^4(t)$ can be expressed by these three functions. Now, the differential equations we need to solve become
\begin{eqnarray}\label{cG relation 1}
    \Lambda(t) &=& K_XX - K -3ELK_XX - \frac{1}{2H_0}K_{X}L\dot{X},\nonumber\\
    c(t) &=& K_XX - 3ELK_XX + \frac{1}{2H_0}K_{X}L\dot{X},\\
    \bar{M}^3_1(t) &=& \frac{2}{H_0}LK_XX,\nonumber
\end{eqnarray}
where $E$ denotes the dimensionless Hubble parameter $H/H_0$. By using the relation
\begin{equation}
    K_X=\frac{\dot{K}}{\dot{X}},
\end{equation}
we can transform Eq.~\eqref{cG relation 1} into
\begin{eqnarray}\label{cG relation 2}
    \dot{K} &=& -\frac{1}{\bar{M}_1^3}\left(\Lambda+c+K+3H\bar{M}_1^3\right)(\Lambda-c+K),\nonumber\\
    L &=& \frac{H_0\bar{M}_1^3}{\Lambda+c+K+3H\bar{M}_1^3},\\
    \dot{X} &=& -\frac{2}{\bar{M}^3_1}X\left(\Lambda-c+K\right)\nonumber.
\end{eqnarray}
As in the previous sections, $K(t)$ and $X(t)$ can be solved from two of these equations, after which we can find $L(t)$ algebraically. In the monotonic interval of $X(t)$, we can solve for $t=t(X)$ and thus derive $K(X)$ and $L(X)$. A sketch of this reconstruction process is displayed in Figure \ref{fig:fc_cG1}.

Analogous to the $k$-essence, for this model, we can also specify the initial condition of $X$, i.e., $X_i$, to be $-M_*^2H_0^2$.\footnote{The specification of the initial condition of $X$ can be considered as the widely-known scaling relation or scaling degeneracy in the cubic Galileon model: the dynamics of this model will stay invariant after scaling the independent variable of the Lagrangian, $X$, by some constant.} However, $K(t)$ now \textit{cannot} be computed by a simple integral like Eq.~\eqref{k X integral} or Eq.~\eqref{k h integral}, implying that the initial condition for $K$, $K_i$, will influence the shape of the reconstructed $K$. In other words, $K_i$ is now important for the reconstruction result. Although it is difficult to obtain its value directly from observations, we can relate it to a more physical parameter. To achieve this, we first write down the equation of motion of the cubic Galileon with shift symmetry \cite{kinetic_braiding_theory}:
\begin{equation}
    \frac{\rm d}{{\rm d}a}\left(a^3J\right)=0,
\end{equation}
with $a$ being the scale factor and
\begin{equation}\label{J definition}
    J=(-2+6EL)(-X)^{1/2}K_X.
\end{equation}
The solution to this equation is
\begin{equation}\label{cG EoM1}
    (-2+6EL)(-X)^{1/2}K_X=\frac{J_0}{a^3},
\end{equation}
where $J_0$ is a nonzero constant. When $J_0=0$, the nontrivial solution to Eq.~\eqref{cG EoM1} is called the tracker solution, or the attractor \cite{kinetic_braiding_theory}, which we will discuss in detail below. If $J_0\neq0$, the solution is said to be off the attractor, and the parameter $J_0$ measures how much the solution deviates from a tracker solution. From Eq.~\eqref{cG relation 0}, the background density of dark energy can now be written as
\begin{equation}\label{cG density}
    \rho_{\rm DE}=c+\Lambda=2K_XX-K-6ELK_XX=-\sqrt{-X}\frac{J_0}{a^3}-K.
\end{equation}
Because we have fixed $X_i=-M_*^2H_0^2$, the initial condition of $K$ can be determined by
\begin{equation}
    K_i=-\rho_{\rm DE}(t_i)-M_*H_0\frac{J_0}{a_i^3},
\end{equation}
where $a_i=a(t_i)$. Evidently, assigning the value of $K_i$ is equivalent to specifying $J_0$ and the dark energy density at time $t_i$.

\begin{figure}[t]
\begin{center}
\includegraphics[width=1\textwidth]{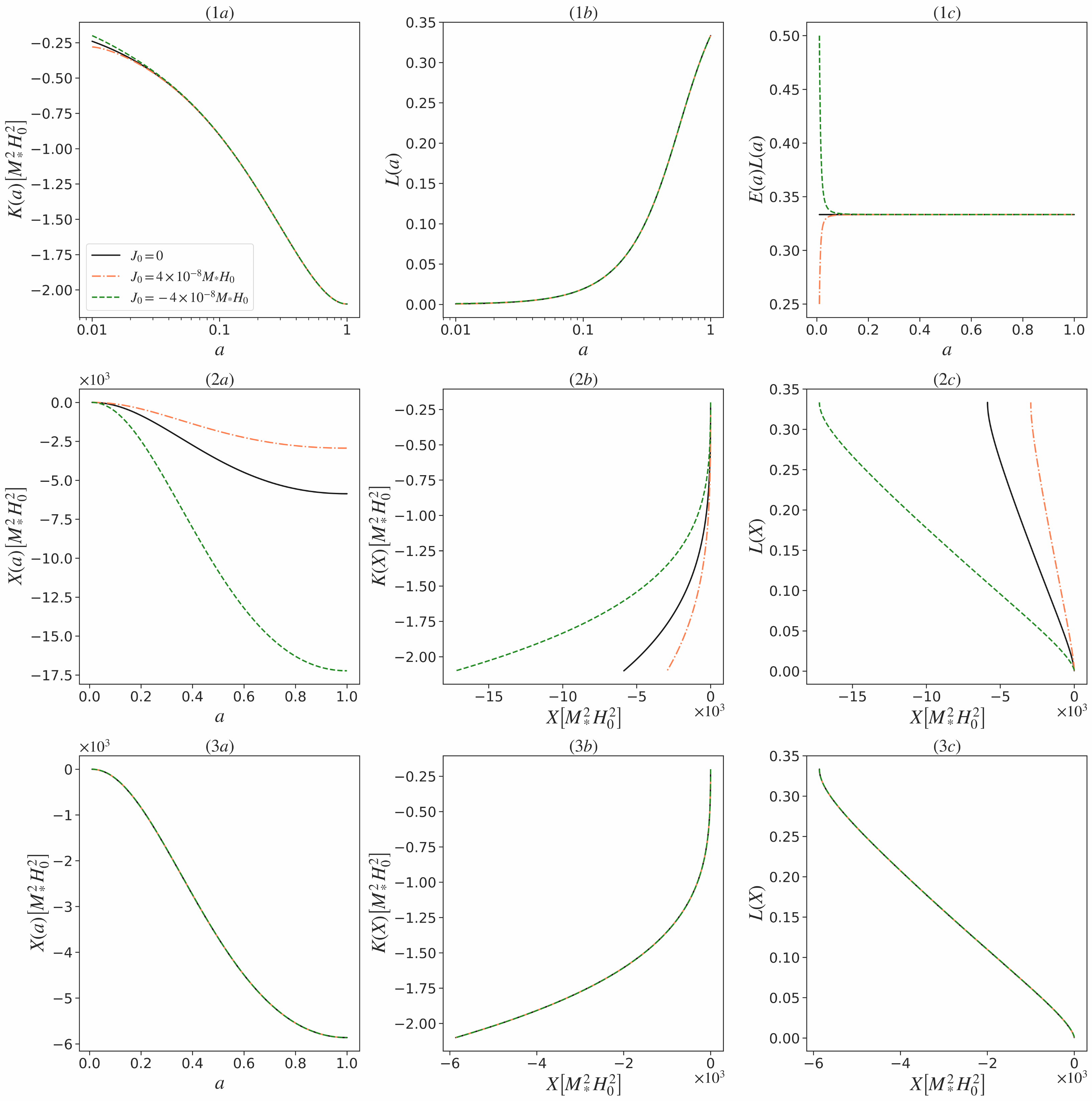}
\caption{The effect of the value of $J_0$ (and thus the initial condition of $K$) on the reconstruction result. Panels (1a)--(1c) depict the evolution of $K(a)$, $L(a)$, and $E(a)L(a)$, which are \textit{time}-dependent quantities specified by the EFT parameters. The time evolution of $X(a)$ (2a) can also be solved in this way, and from all these we can obtain, or reconstruct, $K(X)$ and $L(X)$ (2b and 2c). In all panels we have displayed the results corresponding to $J_0=4\times10^{-8}M_*H_0$ (orange dash-dotted line), $0$ (black solid line), and $-4\times10^{-8}M_*H_0$ (green dashed line). Note that in panels (1a) and (1b) the $a$-axes are drawn using logarithmic scale to make the influence of $J_0$ clearer, and $X, K$ are in units of $M_*^2H_0^2$. The third row is the same as the second, but shows the influence of the value of $J_0$ on the evolution of $X(a)$ and the reconstructed $K(X)$ and $L(X)$ when we appropriately redefine the initial condition $X_i$ for each $J_0$ case (see the text for further details). Here, the $K_i$ values corresponding to the dash-dotted, solid, and dashed curves are the same as in the first two rows, though these lines almost overlap each other.}
\label{fig:J0_cG}
\end{center}
\end{figure}

The reconstruction result for $J_0$ equal to $4\times10^{-8}M_*H_0$, $0$, and $-4\times10^{-8}M_*H_0$ is displayed in the first two rows of Figure \ref{fig:J0_cG}. The results shown in this figure correspond to $w_0=-1$, $w_a=-0.2$, $\Omega_m=0.3$, $\Omega_{\rm DE,0}=0.7$, and a parametrization of $\alpha_B$, the function defined in Eq.~\eqref{alpha_basis} and related to $\bar{M}_1^3(t)$, as
\begin{equation}\label{alpha B parametrization 1}
    \alpha_B(a)=c_B\Omega_{\textrm{DE}}(a)=c_B\frac{\rho_{\textrm{DE}}(a)}{3M_*^2H^2(a)},
\end{equation}
with $c_B$ being a constant and set to $0.5$ in the numerical test here. We solve for $K(a)$, $L(a)$ and $X(a)$ in the range that $0.01<a<1$. In panel (1b), we can see that the value of $J_0$ has very little influence on the evolution of $L(a)$. Panels (1a) and (1c) show that an initial small deviation from the $J_0=0$ solution will reduce toward zero as $a$ increases, which can be expected from Eqs.~\eqref{cG density} and~\eqref{cG relation 2}: as $a$ increases, $K(a)$ converges to $-c(a)-\Lambda(a)$, and $L(a)$ converges to $1/3E(a)$. This is connected to the stability of the tracker solution, which we will discuss in detail in the following paragraphs. In the second row, we see the evolution of $X(a)$ is very different for different $J_0$, leading to very different reconstructed $K(X)$ and $L(X)$. However, as the $K(a)$ corresponding to different $J_0$ converge to the tracker, according to the third equation of Eq.~\eqref{cG relation 2}, $\dot{X}/X$, and thus the increasing rates of $X(a)$, for different $J_0$, will finally become nearly identical. This feature can help us scale the evolution of $X(a)$ to make them converge to a common solution, just like the behavior of $K(a)$ with different initial conditions (or different $J_0$ values) in panel (1a). To achieve this, from the third equation in Eq.~\eqref{cG relation 2}, we can modify the initial condition as
\begin{equation}
\begin{aligned}
     X_i&=-M_*^2H_0^2\frac{X(a=1)\Big|_{X_i=-M_*^2H_0^2,\;J_0=0}}{X(a=1)\Big|_{X_i=-M_*^2H_0^2,\;J_0}}\\&=-M_*^2H_0^2\exp{\left[\int_{a_i}^1\left(\frac{4c(a)}{\bar{M}^3_1(a)}+\frac{2}{\bar{M}^3_1(a)}\left(\Lambda(a)-c(a)+K(a)\right)\right)\frac{\textrm{d}a}{aH(a)}\right]}.
\end{aligned}
\end{equation}
Panels (3a)--(3c) of Figure \ref{fig:J0_cG} show the final results: the corresponding $K_i$ for the dashed-dotted, solid, and dashed lines are identical to the first two rows of this figure, but now the reconstruction results for different $J_0$ are very close, though not exactly identical, to each other. As we mentioned above, this result is related to the property of the tracker solution, and thus indicates that there is further motive to focus on this solution in the reconstruction of this type of theories.

Now, we focus our discussion on the case where $J_0=0$, i.e., the tracker solution. This class of solutions broadly exists in both kinetic gravity braiding \cite{kinetic_braiding_theory} and generalized Galileon \cite{tracker_linear, tracker_power} models, which are independent of $\phi$. The solution can be specified as
\begin{equation}\label{tracker solution}
    E(t)L(X(t))=\frac{1}{3},
\end{equation}
in which the evolution of $X(t)$ by some means tracks the (dimensionless) Hubble parameter. From this solution, we can determine the $L(t)$ by
\begin{equation}\label{tracker L}
    L(t)=\frac{1}{3E(t)}.
\end{equation}
Then from Eq.~\eqref{cG density}, we can derive the evolution of $K$. That is
\begin{equation}\label{tracker K}
\begin{aligned}
    K(t)&=-\rho_{\rm DE}(t)=-c(t)-\Lambda(t)\\&=-3M^2_*H^2_0\Omega_{\rm DE}(t)=3M^2_*H^2_0\left[\Omega^0_ma^{-3}(t)-E^2(t)\right].
\end{aligned}
\end{equation}
Substituting Eq.~\eqref{tracker K} into the last equation of Eq.~\eqref{cG relation 2}, integrating, and using $X_i=-M_*^2H_0^2$, the evolution of $X(t)$ can be computed as
\begin{equation}\label{tracker X}
    X(t)=-M_*^2H_0^2\exp{\left[\int^t_{t_i}\frac{4c(t')}{\bar{M}^3_1(t')}\mathrm{d} t'\right]}.
\end{equation}

\begin{table}[t]
    \centering
    \caption{Parameter sets of four shift-symmetric generalized cubic Galileon models.} 
    \label{table:par_cG}
    \begin{tabularx}{\textwidth}{X<{\centering}|m{90pt}<{\centering} X<{\centering}X<{\centering}X<{\centering}}
        \hline
        Model & Parametrization & $w_0$ & $w_a$ & $c_B$ \\
        \hline
        GI & CPL and Eq.~\eqref{alpha B parametrization 1} & $-1.0$ & $-0.2$ & $0.5$ \\
        GII & CPL and Eq.~\eqref{alpha B parametrization 1} & $-1.0$ & $-0.2$ & $-0.5$ \\
        GIII & CPL and Eq.~\eqref{alpha B parametrization 1} & $-0.8$ & $-0.2$ & $0.5$ \\
        GIV & CPL and Eq.~\eqref{alpha B parametrization 1} & $-0.8$ & $-0.2$ & $-0.5$ \\
        GV & CPL and Eq.~\eqref{alpha B parametrization 2} & $-0.8$ & $-0.4$ & $-0.5$ \\
        GVI & CPL and Eq.~\eqref{alpha B parametrization 2} & $-0.8$ & $-0.4$ & $0.5$ \\
        \hline
    \end{tabularx}
\end{table}

Equations \eqref{tracker L}, \eqref{tracker K}, and \eqref{tracker X} establish a simpler reconstruction framework compared to Eq.~\eqref{cG relation 2}. To apply them, we need to ensure that the evolution of dark energy is on the tracker solution, or it has converged to the tracker at an early enough time. In other words, the tracker solution should be stable. \textcolor{black}{(We emphasize that the ``stability" we are considering here denotes the convergence to the tracker solution for a small perturbation off the tracker, which is different from what we have discussed in the last section, where the stability means the evolution of physical quantities like the density perturbation or scalar field won't be ill-behaved.)} The stability condition has been researched where $K(X)$ and $G_3(X)$ are linear or power-law functions of $X$ \cite{tracker_linear, tracker_power}, and for covariant cubic Galileon some studies claim that the CMB angular power spectrum will differ too much from observations if the solution fails to converge onto the tracker early enough \cite{tracker_analysis}. For the general case we are considering, we first define
\begin{equation}
    r=EL(X),
\end{equation}
and thus, a solution converging to the tracker is equivalent to $r\rightarrow\frac{1}{3}$. The second Friedmann equation can be written as
\begin{equation}
    2M^2_*\dot{H}+3M^2_*H^2=-M_*^2H^2\Omega_r-K(X)-\frac{L(X)}{H_0}K_X(X)\dot{X},
\end{equation}
where $\Omega_r=\rho_r/(3M_*^2H^2)$ is the $a$-dependent density parameter for radiation. Together with the field equation of motion (which is obtained by expanding Eq.~\eqref{cG EoM1})
\begin{equation}
\begin{aligned}
    &3H\left[-2(-X)^{1/2}K_X(X)+6E(-X)^{1/2}L(X)K_X(X)\right]+(-X)^{-1/2}K_X(X)\dot{X}\\&-2(-X)^{1/2}K_{XX}(X)\dot{X}+6\dot{E}(-X)^{1/2}L(X)K_X(X)- 3E(-X)^{-1/2}L(X)K_X(X)\dot{X}\\&+6E(-X)^{1/2}L_X(X)K_X(X)\dot{X}+6E(-X)^{1/2}L(X)K_{XX}(X)\dot{X}=0,
\end{aligned}
\end{equation}
we can express $\dot{E}$ and $\dot{X}$ as functions of $E$ and $X$ and derive that
\begin{equation}
    \frac{r'}{r}=\frac{L_X}{EL}\dot{X}+\frac{\dot{E}}{E^2}=-(3r-1)\frac{\Delta_n}{\Delta_d},
\end{equation}
where a prime denotes derivative with respect to $N=\ln{a}$, and
\begin{equation}
\begin{aligned}
    \Delta_n=&-6E\frac{X}{M^2_*H^2_0}L^2(X)K^2_X(X)+2X\big(E^2(3+\Omega_r)+\frac{K(X)}{M^2_*H^2_0}\big)L(X)K_{XX}(X)\\&+\big[\big(E^2(3+\Omega_r)+\frac{K(X)}{M^2_*H^2_0}\big)L(X)+12E^2XL_X(X)\big]K_X(X),\\
    \Delta_d=&2E^2L(X)\big[-3\frac{X}{M^2_*H^2_0}L^2(X)K^2_{X}(X)+2X\big(3EL(X)-1\big)K_{XX}\\&+\big(3EL(X)-1+6EXL_X(X)\big)K_X(X)\big].
\end{aligned}
\end{equation}
We know that the tracker solution corresponds to the fixed point $r_t=\frac{1}{3}$, where the subscript $t$ stands for ``tracker". If we add a small perturbation $\delta{r}$ to the fixed point, that is $r=\frac{1}{3}+\delta r$, then up to the linear order in $\delta{r}$ we have
\begin{equation}
    \delta r'=-\frac{\Delta_{n,t}}{\Delta_{d,t}}\delta r,
\end{equation}
with
\begin{equation}
\begin{aligned}
    \Delta_{n,t}=&-\frac{2X_t}{M^2_*H^2_0}L(X_t)K^2_X(X_t)+12X_tE_t^2L_X(X_t)K_X(X_t)\\&+\Big(2X_tK_{XX}(X_t)+K_X(X_t)\Big)\Big(E_t\big(1+\frac{1}{3}\Omega_r\big)+\frac{K(X_t)}{M^2_*H^2_0}L(X_t)\Big),\\
    \Delta_{d,t}=&-\frac{2X_t}{3M^2_*H^2_0}L(X_t)K^2_{X}(X_t)+4E_t^2X_tL_X(X_t)K_X(X_t),
\end{aligned}
\end{equation}
where $X_t(t)$ denotes the value of $X(t)$ on the tracker solution. Thus, the tracker solution is (locally) stable as long as $\Delta_{n,t}/\Delta_{d,t}>0$. Also, if $\Delta_{n,t}/\Delta_{d,t}>n>0$, we have $\frac{\rm d}{{\rm d}\ln{a}}\ln{|\delta r|}<-n$, and thus $|\delta r|<|\delta r_i|a^{-n}$, which means the solution will converge to the tracker solution faster than $a^{-n}$. We also note here that by using the EFT parameters, the stability condition can be rewritten as
\begin{equation}\label{stability condition}
\begin{aligned}
    \frac{\Delta_{n,t}}{\Delta_{d,t}}=3-\frac{1+\frac{1}{3}\Omega_r-\Omega_{DE}}{H[\Omega_m+\frac{4}{3}\Omega_r]} \left(2\frac{c}{\bar{M}_1^3}-\frac{\dot{\bar{M}}_1^3}{\bar{M}_1^3}-\frac{\dot{H}}{H} \right)>0.
\end{aligned}
\end{equation}

\begin{figure}[t]
\begin{center}
\includegraphics[width=1\textwidth]{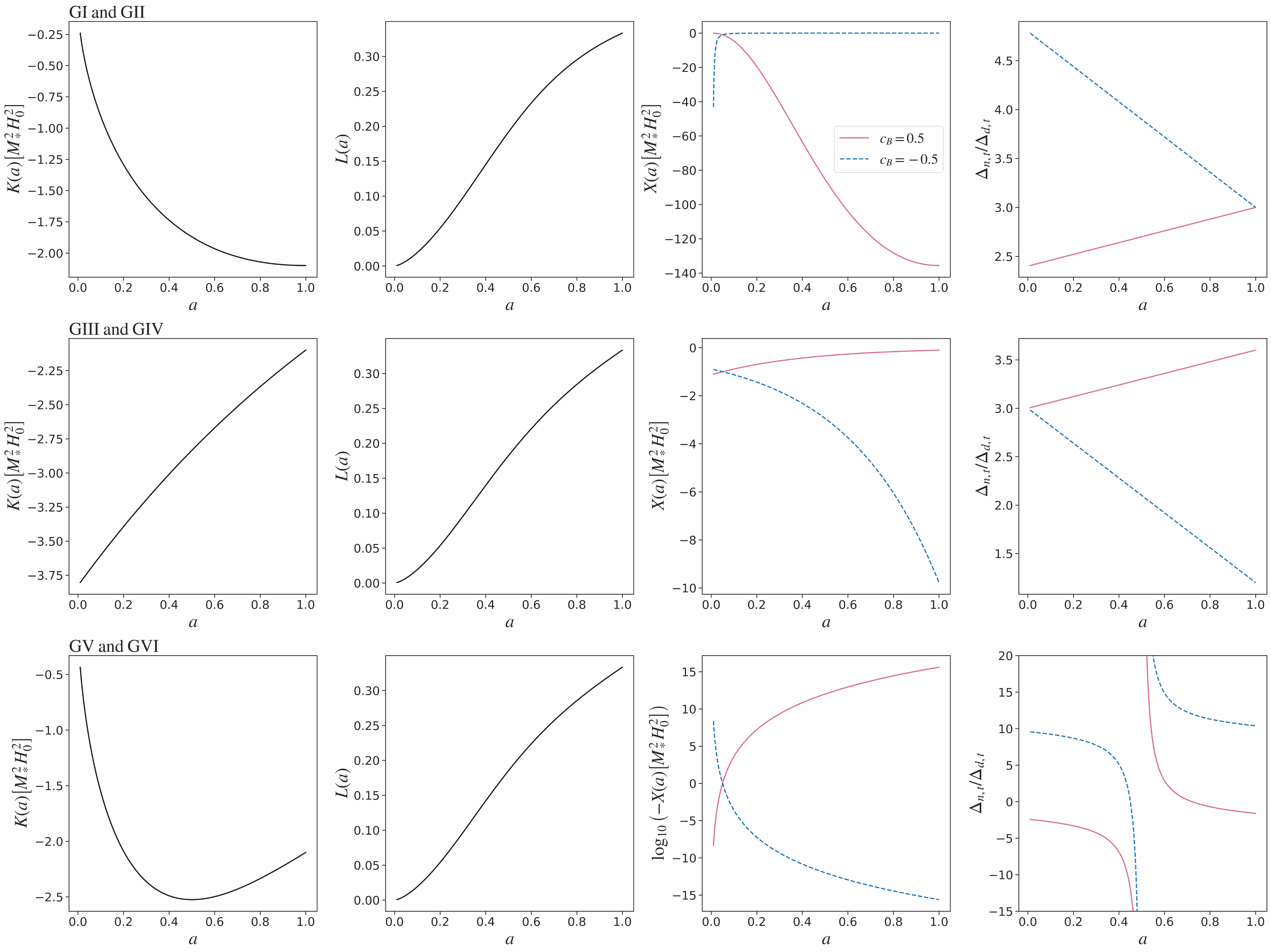}
\caption{An example of reconstruction for the shift-symmetric generalized cubic Galileon models, assuming they are on the tracker solution. \textit{The first row}: the first two panels show the evolution of $K(a)$ and $L(a)$, determined by the chosen EFT parameters, in models GI and GII, where the background dark energy densities have identical evolution, leading to identical $K(a)$ and $L(a)$. Note that $K$, $L$, and $X$ are in units of $M_*^2H_0^2$, $1$, and $M_*^2H_0^2$, respectively. The third panel displays the evolution of $X(a)$, which is also determined by the chosen EFT parameters, in unit of $M_*^2H_0^2$, where the purple solid and blue dashed lines represent, respectively, the models with a positive (GI) and negative (GII) $c_B$. The last panel shows the evolution of $\Delta_{n,t}/\Delta_{d,t}$, which appears in the LHS of the stability condition, Eq.~\eqref{stability condition}; the color is the same as in the third panel. \textit{The second row} and \textit{the third row}: the evolution of $K(a)$, $L(a)$, $X(a)$, and the stability condition in models GIII \& GIV and GV \& GVI, respectively, with the same arrangement, except that in the third row, the y-axis of the third panel uses logarithmic scale.} 
\label{fig:nu_cG}
\end{center}
\end{figure}

\begin{figure}[t]
\begin{center}
\includegraphics[width=1\textwidth]{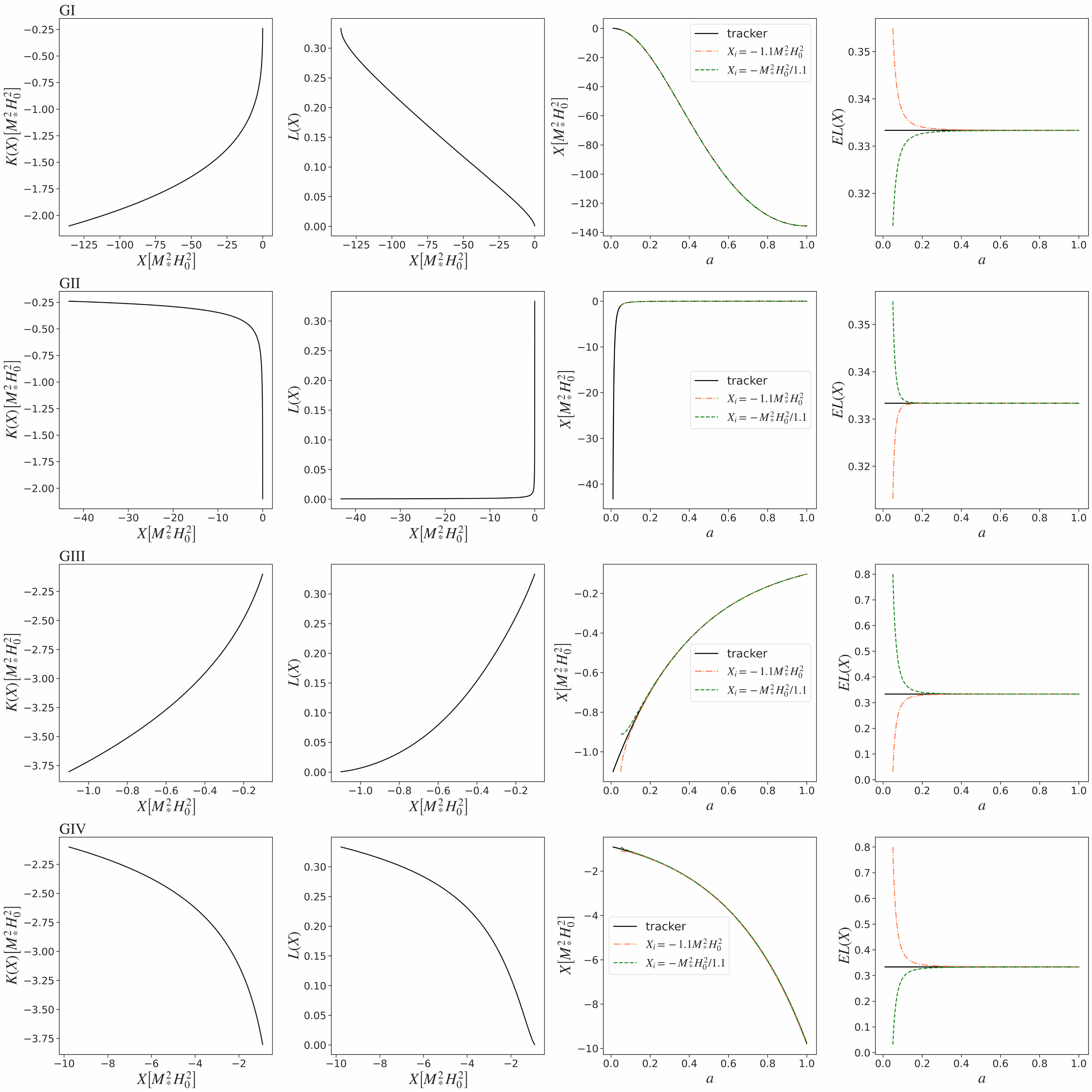}
\caption{The reconstructed $K(X)$ and $L(X)$ of the models shown in Fig.~\ref{fig:nu_cG} and the ``convergence tests'' of the perturbation off the tracker for models without phantom crossing. \textit{The first row}: In the first two panels, the reconstructed $K(X)$ and $L(X)$ for model GI are obtained from the evolution of $X(a)$, $K(a)$ and $L(a)$ in Fig.~\ref{fig:nu_cG}. The third panel shows the evolution of the tracker solution (black solid line), as well as solutions with $X_i=X(a=0.05)=-1.1M_*^2H_0^2$ (orange dashed-dotted line) and $X_i=X(a=0.05)=-M_*^2H_0^2/1.1$ (green dashed line). The last panel exhibits the evolution of $E(a)L(a)$ of these solutions, whose convergence or divergence relative to the tracker presents the stability of the tracker solution. $K$, $L$, and $X$ are in units of $M_*^2H_0^2$, $1$, and $M_*^2H_0^2$, respectively. \textit{The second, third}, and \textit{fourth rows} correspond to model GII, GIII, and GIV, whose configurations are identical to the first row.}
\label{fig:con_cG1}
\end{center}
\end{figure}

Just as for quintessence and $k$-essence, before presenting the numerical results of the simplified reconstruction method, we first briefly discuss how to choose a proper form of the EFT parameters to ensure the validity of the reconstruction framework. When dark energy is on the tracker, $L(t)$ monotonically increases as long as the dimensionless Hubble parameter $E(t)$ monotonically decreases, which is a reasonable assumption. If $X(t)$ is not monotonic, the reconstructed $L(X)$ will become multi-valued. Consequently, $X(t)$ must be monotonic. This indicates that the formula in the integral of Eq.~\eqref{tracker X}, $4c(t)/\bar{M}_1^3(t)$, should not change the sign during the period of time we are interested in. Following the same analysis as in the previous sections, the value of $c(t)$ will cross zero once the equation of state of dark energy, $w_{\textrm{DE}}$, crosses $-1$. That subsequently implies that the value of $\bar{M}_1^3(t)$ will cross zero during the phantom crossing event.


The zero crossing of $\bar{M}^3_1$ leads to $\dot{\bar{M}}^3_1/\bar{M}^3_1= {\rm d}\ln{\bar{M}^3_1}/{\textrm{d}t}$ being unbounded. Before and after the zero crossing, the sign of $\bar{M}^3_1$ will change, and the sign of $\dot{\bar{M}}^3_1$ remains unchanged. That indicates that on one side of the zero point, $\dot{\bar{M}}^3_1/\bar{M}^3_1$ will diverge to negative infinity, which thereby implies the tracker solution is unstable at this point when compared to Eq.~\eqref{stability condition}.\footnote{Because the ratio $c(t)/\bar{M}_1^3(t)$ appears in the integral which determines the evolution of $X(t)$, it's reasonable to assume it remains finite to ensure $X(t)$ to be well-behaved.} This instability can actually also be observed from a different aspect. On the tracker, one has $J=0$, where $J$ has been defined in Eq.~\eqref{J definition}. If the solution starts off the tracker, then $J\neq0$ while $Ja^3$ stays constant by virtue of the equation of motion Eq.~\eqref{cG EoM1}. Assume that the solution is able to continuously get closer to the tracker solution, which is equivalent to stability against (small) perturbations. Assume also that $w_{\rm DE}$ crosses $-1$. Because on the tracker $\rho_{\rm DE}=-K$, at the phantom crossing $K$ will reach an extreme value so that $K_X=0$, where for $Ja^3$ to remain a finite constant we must have $-2+6EL\rightarrow\infty$. As a result, whenever the dark energy undergoes a phantom-crossing event, the tracker solution will become unstable, making Eqs.~\eqref{tracker L} --~\eqref{tracker X} invalid.

Our discussion of stability above is focused on the convergence of the background equations of motion toward the tracker solution. Now, assuming the background evolution follows the tracker, we turn to analyze the \textcolor{black}{gradient} stability of dark energy density perturbations. For cubic Galileon models, the sound speed of dark energy can be written as \cite{EFT_theory2}
\begin{equation}\label{cG sound speed}
    c_s^2=\frac{4c-2H\bar{M}^3_1-2\dot{\bar{M}}^3_1-\left(\bar{M}^3_1\right)^2/M_*^2}{4c+8M^4_2+3\left(\bar{M}^3_1\right)^2/M_*^2}.
\end{equation}
On the tracker, Eq.~\eqref{cG relation 0} and Eqs.~\eqref{tracker L} --~\eqref{tracker X} lead to
\begin{equation}
\begin{aligned}
    c+2M_2^4&=-6EX^2L_XK_X-2XK_X\\&=-6E\left(\frac{X}{\dot{X}}\right)^2\dot{L}\dot{K}-2\frac{X}{\dot{X}}\dot{K}=-3H\left(1-\frac{1}{4}\frac{\bar{M}^3_1}{c}\frac{\dot{H}}{H}\right)\bar{M}^3_1.
\end{aligned}
\end{equation}
Substituting this relation into Eq.~\eqref{cG sound speed}, we obtain
\begin{equation}
    c_s^2=\frac{4\left(c/\bar{M}^3_1\right)-2H-2\left(\dot{\bar{M}}^3_1/\bar{M}^3_1\right)-\bar{M}^3_1/M_*^2}{3\bar{M}^3_1/M_*^2-12H\left(1-\frac{1}{4}\frac{\bar{M}^3_1}{c}\frac{\dot{H}}{H}\right)}.
\end{equation}
Thus, for the same reason as the last paragraph, the sound speed will become imaginative on one side of the phantom crossing, leading to an instability in the evolution of the DE density perturbation.

\begin{figure}[t]
\begin{center}
\includegraphics[width=1.0\textwidth]{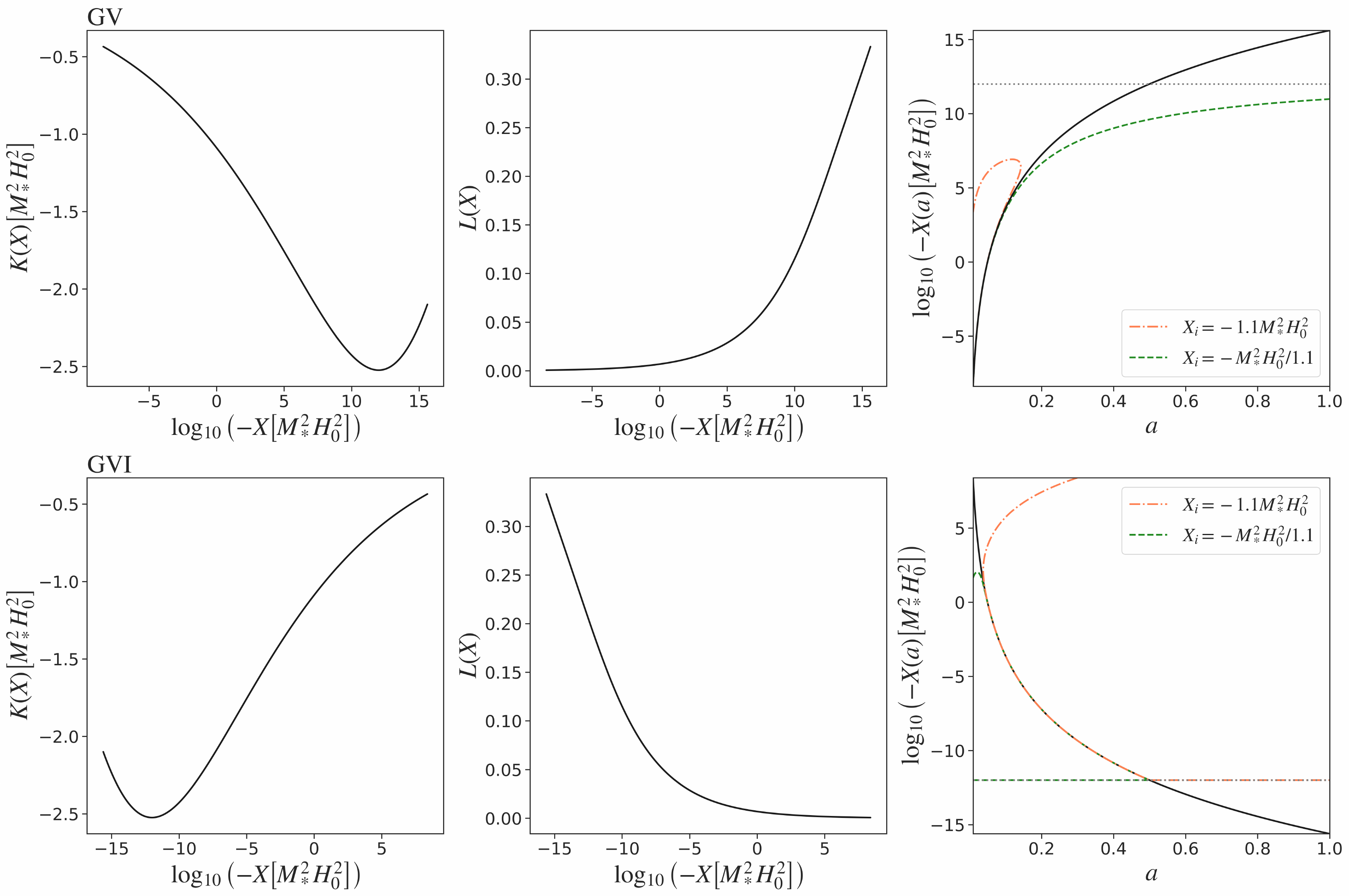}
\caption{The reconstructed $K(X)$ and $L(X)$ and the ``convergence tests'' of the perturbation off the tracker for models with phantom crossing. \textit{The first row}: In the first two panels, the reconstruction results of $K(X)$ and $L(X)$ for model GV are obtained from the evolution of $X(a)$, $K(a)$ and $L(a)$. The last panel shows the evolution of the tracker solution (the black solid line), as well as contours with $X_i=X(a=0.05)=-1.1M_*^2H_0^2$ (orange dashed-dotted line) and $X_i=X(a=0.05)=-M_*^2H_0^2/1.1$ (the green dashed line) whose convergence or divergence relative to the tracker present the stability of the tracker solution. The horizontal gray line signifies the value of $X$ when $w_{\textrm{DE}}$ crosses the $-1$. See the description in the text for more details. The $K$, $L$, and $X$ are drawn in units of $M_*^2H_0^2$, $1$, and $M_*^2H_0^2$, respectively. \textit{The second row} corresponds to model GVI, whose configuration is identical to the first row.}
\label{fig:con_cG2}
\end{center}
\end{figure}

Finally, we present several concrete examples applying the reconstruction framework that we have discussed in this section. We first pay our attention to models \textit{without} a phantom-crossing event, with the evolution of $w_{\textrm{DE}}$ fixed by CPL parametrization and the evolution of $\bar{M}_1^3$ fixed by Eq.~\eqref{alpha B parametrization 1}. The parameter sets are exhibited in the first four rows of Table \ref{table:par_cG}. In the first two rows of Figure \ref{fig:nu_cG}, we present the solution of Eq.~\eqref{cG relation 2} and the left-hand side of the stability condition~\eqref{stability condition} for all four models, and despite solving these functions for $a\in[0.01, 1]$, we fix $X_i=X(a=0.05)=-M_*^2H_0^2$ for future convenience. From this figure, $\Delta_{n,t}/\Delta_{d,t}$ is always positive for every model, leading to the conclusion that these tracker solutions are all stable. 

To verify this, we show the reconstructed $K(X)$ and $L(X)$ in the first two columns of Figure \ref{fig:con_cG1}, and use them along with Eq.~\eqref{cG EoM1} to study the evolution of background perturbations. Similar to the discussion in the last section, the framework is restricted to reconstructing only one part of the full Lagrangian. Consequently, if we start solving the perturbations when $a=1$, or choose the initial value too far away from the tracker, the evolution of $X(a)$ at early times would exceed the range we reconstruct. We therefore take $a=0.05$ as the initial point of the perturbed solutions, and fix $X_i$ to be $-1.1M_*^2H_0^2$ and $-M_*^2H_0^2/1.1$ for both of them. From the last two columns of this figure, the solutions off the tracker will rapidly converge to the tracker solution for all these four models. Additionally, the values of $\Delta_{n,t}/\Delta_{d,t}$ indeed reflect the convergence speed. Take model GI and GII as an example, where the latter's $\Delta_{n,t}/\Delta_{d,t}$ is greater than the former's. Qualitatively, from the last column of the first two rows of Figure \ref{fig:con_cG1}, the quantity $E(a)L(a)$ of model GI does converge to $\frac{1}{3}$ more slowly. Quantitatively, for model GII, the relative difference between $E(a)L(a)$ and $\frac{1}{3}$ becomes less than $10^{-4}$ when $a\geq0.17$, while for model GI, it happens only when $a\geq0.43$.

To illustrate the instability of models \textit{with} phantom crossing, we implement our reconstruction framework on model V and model VI, whose parameter sets are given in the last two rows of Table \ref{table:par_cG}. As in this case the sign of $\bar{M}^3_1(a)$ changes when $w_{\textrm{DE}}$ crosses $-1$, we use the parametrization
\begin{equation}\label{alpha B parametrization 2}
    \alpha_B(a)=c_B(1+w_{\textrm{DE}})\Omega_{\textrm{DE}}(a)=c_B(1+w_{\textrm{DE}})\frac{\rho_{\textrm{DE}}(a)}{3M_*^2H^2(a)},
\end{equation}
with $c_B$ again a constant. As before, we solve for $K(a)$, $L(a)$ and $X(a)$ for $a\in[0.01, 1]$, fixing $X_i=X(a=0.05)=-M_*^2H_0^2$. The solutions of Eq.~\eqref{cG relation 2} for these models are presented in the last row of Figure \ref{fig:nu_cG}. From the last panel of this row, we see that these two models both diverge before $a=0.5$, where the phantom crossing happens, consistent with what we have discussed before. We have displayed the reconstructed $K(X)$ and $L(X)$ in the first two columns of Figure \ref{fig:con_cG2}. However, because of the instability near the phantom crossing, most of the algebraic equation solvers and differential equation solvers break down before that point. Therefore, here we employ another method to present the perturbations. The equation of motion for generalized cubic Galileon with shift symmetry, i.e., Eqs.~\eqref{J definition} and \eqref{cG EoM1}, can be rewritten as
\begin{equation}\label{cG EoM2}
    a^3J=J_0.
\end{equation}
If we interpret $a^3J$ as a function of $a$ and $X$, this equation corresponds to a contour on the $a$--$X$ plane for every specific $J_0$. Subsequently, we actually present the contours corresponding to perturbations with $X_i=X(a=0.05)$ being $-1.1M_*^2H_0^2$ and $-M_*^2H_0^2/1.1$ in the last column of Figure \ref{fig:con_cG2}. In these panels, we also display $X_c$, the value of $X(a)$ at the phantom crossing, on which $K_X(X_c)=0$ and therefore $a^3J=0$. As a result, contours for a non-zero $J_0$ will never cross this line. For model GV, $\Delta_{n,t}/\Delta_{d,t}$ is always negative at $a\leq0.5$, and the green dashed and orange dash-dotted curves indeed evolve away from the tracker. More interestingly, there is a ``turnaround" in the orange curve, meaning this solution will actually not exist in the future. The off-the-tracker perturbations for model GVI behave similarly. It seems that there are spikes in the green and the orange contours at $a=0.5$. These contours actually turn around smoothly, but the solutions first converge to the tracker very closely and only evolve away from it very near the phantom crossing, making the turnarounds look acute. Apart from that, there is a turnaround for the orange curve in the upper left part of this panel, leading to the fact that this solution will not even exist in the past. These results verify the conclusion that it is impossible to have a stable tracker solution and a valid reconstruction framework Eq.~\eqref{tracker L} --~\eqref{tracker X} for a shift-symmetric cubic Galileon once there is a phantom-crossing event.







\section{Discussions and conclusions}
\label{sec:discuss}

We presented a general method to reconstruct DE theories fully non-linearly. This is based on the fact that, given a subclass of DE theory, a single Lagrangian governs the background, linear and non-linear dynamics, and these aspects are correlated rather than independent of each other. The implementation of this method only requires observations of the cosmic expansion history and certain linear perturbation quantities, which can be described by the effective field theory of dark energy parameterization. We connect the EFTofDE parameters, which are a series of time-dependent functions, to (sub)classes of Horndeski theory via some equations. Through solving these algebraic and differential equations, we determine the values of various functions in the theoretical Lagrangian along the path $(\phi(t), X(t))$ that is followed by the background evolution. These reconstructions would then allow us to study the fully non-linear evolution in such theories. We have illustrated that this method can be implemented on a number of dark energy models, which are general enough and have been extensively researched in the literature.

The idea underlying the method is similar to that behind modified gravity tomography \citep{MG_tomography,MG_tomography2}, where a paramterization of the scalar field mass $m$ and coupling strength $\beta$ as time functions, which are needed to specify the background and linear perturbation evolution in thin-shell screening models, also fix the full Lagrangian which can be used to study the non-linear dynamics in both time and space \citep{MG_tomography_sim1,MG_tomography_sim2}. Together, these will enable the following strategy for studying dark energy and modified gravity models, especially those with screening mechanisms, in a systematic and comprehensive way: 
\begin{enumerate}[topsep=3pt, itemsep=3pt, parsep=0pt]
    \item identify categories of background expansion history, such as $\Lambda$CDM, $w_0$-$w_a$ without phantom crossing, and $w_0$-$w_a$ with phantom crossing;
    \item for each expansion history, identify the classes of DE and MG theories allowed;
    \item reconstruct the corresponding Lagrangians;
    \item explore the space of reconstructed theories using analytical and numerical approaches, including simulations, to constrain the original parameters using cosmological probes.
\end{enumerate}

Our reconstruction framework cannot produce the full functional forms of $K(\phi, X)$, $G_3(\phi, X)$, and $G_4(\phi)$: indeed often the reconstructed result has slightly different \textit{ranges} of $\phi$ and $X$ from what we need to study the non-linear evolution. For functions in the Lagrangian depending on $X$ or $\phi$ only, e.g., $G_3(X)$ and $K(X)$, they are reconstructed in a 1D interval of $X$ or $\phi$. If the Lagrangian allows a separation of variables, as considered in Section \ref{sec:kessence}, we can reconstruct a 2D area on the $\phi$--$X$ plane. In all cases, a non-linear study may need the values of these functions slightly outside the reconstructed ranges of $\phi$ and/or $X$, but the missing information is expected to be insignificant. Indeed, we notice that in most DE models that do not involve thin-shell screening, even on the small, highly non-linear, scales, the perturbations of $X$ and $\phi$ are generally small and negligible. In this sense the practicality of our reconstruction method in non-linear studies is not compromised. 

The complete reconstruction of DE models requires solving differential equations, necessitating the initial conditions for $K$, $G_3$ and $G_4$, which would naively appear to be observationally challenging to get. Fortunately, we demonstrated that for the quintessence, scalar-tensor and $k$-essence theories considered in this paper, we actually do not need to specify the initial conditions, as they can either be fixed to specific values via a field redefinition or be canceled out in the final results. For the shift-symmetric cubic Galileon discussed in Section \ref{sec:galileon}, the situation is more complicated: we found that the initial condition $K_i$, which is linked to the constant $J_0$ in the equation of motion, does influence the reconstruction results. However, the tracker solution can simplify the reconstruction process, by converting the differential equation for $K$ into an algebraic one. We studied the validity of this simplification by discussing the convergence property around the tracker solution. We demonstrated that for most situations when the equation of state $w_{\textrm{DE}}$ does not cross $-1$, an initially small perturbation rapidly converges to the tracker. As a result, we consider it is a reasonable assumption that our Universe, at least for the period that we are interested in, is always on a tracker. However, when the model does possess a phantom crossing, the stability condition is violated near this event, and the solutions that start off the tracker would either move away from the tracker or stop existing after some point, invalidating the simplified reconstruction method.

In this paper, we have briefly discussed the influence of the gradient instability, i.e., when the sound speed square $c_s^2$ becomes negative, on the reconstruction results in Sections \ref{sec:kessence} and \ref{sec:galileon}, but left other instabilities, such as the ghost or tachyon instability, unexamined, since a detailed discussion on these topics is beyond the scope of this paper. Nevertheless, we point out that the stability condition for dark energy's density perturbation can be directly determined through the observed EFT parameters. Provided the observed evolution of these parameters permits a healthy evolution of the density perturbations, the reconstructed DE theories should correspondingly exhibit well-behaved properties, at least on the linear scales.

For clarity of the presentation, in this paper, we always utilize the CPL parametrization to specify the background evolution of the dark energy with some additional assumptions, such as a constant sound speed or the EFT parameters being proportional to the dark energy density ratio. Nonetheless, we note here that these assumptions are unnecessary. The present and future observations of supernova, baryonic acoustic oscillation, weak lensing and perhaps the direct detection of the redshift drift, along with some non-parametric methods, such as the Gaussian process regression or other machine learning algorithms, will enable a precise measurement of the expansion history and linear structure growth, and thus the evolution of the EFT parameters, to a relatively high redshift, thereby offering a solid foundation for applying this reconstruction method.

\textcolor{black}{Throughout this paper we have focused on the first five EFT parameters, which provide us at most five independent reconstruction equations (Eq.~\eqref{EFT relation}). This implies that the number of reconstructible free functions in the dark energy Lagrangian is fundamentally limited to five or fewer, according to the model we consider. For instance, in the case of $k$-essence, there are only three relevant EFT parameters, and subsequently, we can reconstruct up to three univariate functions. Specifically, for a Lagrangian of the form $K(\phi, X)=V(\phi)+f(\phi)X+q(\phi)X^2$, field redefinition allows us to determine $\phi(t)$, thereby enabling the reconstruction of $V(\phi)$, $f(\phi)$, and $q(\phi)$. In contrast, for separable models like $K(\phi, X)=q(\phi)h(X)$, one EFT equation must be allocated to determine the form of $\phi(t)$ and $X(t)$, reducing the number of reconstructible functions to two.} \textcolor{black}{Going beyond this, as we discuss below, for more complicated cases of $K(\phi,X)$ it may not be necessary to reconstruct the individual uni-variate functions such as $q(\phi)$ and $h(X)$, but instead we can treat $K(\phi,X)$ as a 2D function, the values of which and whose derivatives along the 1D curve in the $(\phi,X)$ plane around the background solution can be obtained numerically, offering enough information for cosmological analyses.}


We also note here that though in this paper we only utilized 5 EFT parameters and restricted our discussion to models like quintessence, scalar-tensor theory, $k$-essence, and shift-symmetric generalized cubic Galileon in this paper, our reconstruction method is more general, and it is straightforward to extend the examples above to more EFT parameters and more complicated dark energy or modified gravity models. For instance, we can generalize the methodology described in Section \ref{sec:galileon} to reconstruct models with $K=q(\phi)h(X)$ and $G_3=G_3(X)$ by using 4 EFT parameters, $c(t)$, $\Lambda(t)$, $M_2^4(t)$ and $\bar{M}_1^3(t)$, or we can even include terms in the ellipsis of Eq.~\eqref{EFT action} to reconstruct theories involving the Horndeski $L_5$ Lagrangian. 

In the most general case of generalized Galileon, where $K=K(\phi, X)$ and $G_3=G_3(\phi, X)$, although we can only know the values of $K$ and $G_3$ on a 1D path instead of a 2D region in the $\phi$-$X$ plane (as discussed at the beginning of Section \ref{sec:kessence}), it is still possible to extract information about non-linear dynamics from background and linear scales. \textcolor{black}{The key thing is that the 1D curve is what the background solution of our model universe is on (especially since we presumably use background cosmological observations to fix the EFTofDE parameterisation), and knowing only it rather than the K function in the 2D phi-X space, will not prevent us from studying the full nonlinear evolution along this 1D curve.} As we mentioned before, the spatial perturbations of the scalar field are usually small (it is the spatial derivatives of these perturbations that are non-negligible), and it is indeed a common assumption in simulations of such models that the coefficients of the non-linear terms depend only on the background. Thus, \textcolor{black}{within the scope of cosmological applications,} for every specific model, its non-linear dynamical properties can be retrieved \textcolor{black}{in a ``just enough" fashion} from its background and linear evolution. Once we have reconstructed a DE theory from the EFT parameters, we can apply this to cosmological simulations, as previously stated. The differences between simulations and observations could provide evidence to constrain or rule out this model. 


In conclusion, this reconstruction framework connects the background and linear evolution to the phenomena at the fully non-linear scales, and offers an accessible and effective way to verify different dark energy models. More importantly, it enables us to only focus on DE models that satisfy certain targeted background and linear-perturbation behavior, out of the infinite choices of DE Lagrangians, offering convenience for theoretical and numerical studies. We plan to demonstrate its application in such simulation studies in a series of follow-up works.


\acknowledgments

YG and JW are supported by the National Key R\&D Program of China (2022YFA1602901), the NSFC Grant (No. 12125302), CAS Project for Young Scientists in Basic Research Grant (No. YSBR-062), and the China Manned Space Program Grant (Nos. CMS-CSST-2025-A03 and CMS-CSST-2025-A10). BL is funded by UK STFC through Consolidate Grant ST/X001075/1, and thanks the National Astronomical Observatories of China and Beijing Normal University for their host when part of the work described here was carried out.

\bibliographystyle{JHEP}
\bibliography{biblio.bib}

\providecommand{\href}[2]{#2}\begingroup\raggedright\begin{thebibliography}{100}

\bibitem{dark_energy_supernovae_Riess}
A.G.~{Riess}, A.V.~{Filippenko}, P.~{Challis}, A.~{Clocchiatti}, A.~{Diercks},
  P.M.~{Garnavich} et~al., \emph{{Observational Evidence from Supernovae for an
  Accelerating Universe and a Cosmological Constant}},
  \href{https://doi.org/10.1086/300499}{\emph{Astron. J.} {\bfseries 116}
  (1998) 1009} [\href{https://arxiv.org/abs/astro-ph/9805201}{{\ttfamily
  astro-ph/9805201}}].

\bibitem{dark_energy_supernovae_Perlmutter}
{\scshape Supernova Cosmology Project} collaboration, \emph{{Measurements of
  $\Omega$ and $\Lambda$ from 42 High Redshift Supernovae}},
  \href{https://doi.org/10.1086/307221}{\emph{Astrophys. J.} {\bfseries 517}
  (1999) 565} [\href{https://arxiv.org/abs/astro-ph/9812133}{{\ttfamily
  astro-ph/9812133}}].

\bibitem{quintessence_theory1}
B.~{Ratra} and P.J.E.~{Peebles}, \emph{{Cosmological consequences of a rolling
  homogeneous scalar field}},
  \href{https://doi.org/10.1103/PhysRevD.37.3406}{\emph{Phys. Rev. D}
  {\bfseries 37} (1988) 3406}.

\bibitem{quintessence_theory2}
P.G.~{Ferreira} and M.~{Joyce}, \emph{{Structure Formation with a Self-Tuning
  Scalar Field}},
  \href{https://doi.org/10.1103/PhysRevLett.79.4740}{\emph{Phys. Rev. Lett.}
  {\bfseries 79} (1997) 4740}
  [\href{https://arxiv.org/abs/astro-ph/9707286}{{\ttfamily
  astro-ph/9707286}}].

\bibitem{quintessence_theory3}
E.J.~{Copeland}, A.R.~{Liddle} and D.~{Wands}, \emph{{Exponential potentials
  and cosmological scaling solutions}},
  \href{https://doi.org/10.1103/PhysRevD.57.4686}{\emph{Phys. Rev. D}
  {\bfseries 57} (1998) 4686}
  [\href{https://arxiv.org/abs/gr-qc/9711068}{{\ttfamily gr-qc/9711068}}].

\bibitem{quintessence_theory4}
R.R.~{Caldwell}, R.~{Dave} and P.J.~{Steinhardt}, \emph{{Cosmological Imprint
  of an Energy Component with General Equation of State}},
  \href{https://doi.org/10.1103/PhysRevLett.80.1582}{\emph{Phys. Rev. Lett.}
  {\bfseries 80} (1998) 1582}
  [\href{https://arxiv.org/abs/astro-ph/9708069}{{\ttfamily
  astro-ph/9708069}}].

\bibitem{quintessence_theory5}
I.~{Zlatev}, L.~{Wang} and P.J.~{Steinhardt}, \emph{{Quintessence, Cosmic
  Coincidence, and the Cosmological Constant}},
  \href{https://doi.org/10.1103/PhysRevLett.82.896}{\emph{Phys. Rev. Lett.}
  {\bfseries 82} (1999) 896}
  [\href{https://arxiv.org/abs/astro-ph/9807002}{{\ttfamily
  astro-ph/9807002}}].

\bibitem{FR_theory1}
S.~{Nojiri} and S.D.~{Odintsov}, \emph{{Introduction to Modified Gravity and
  Gravitational Alternative for Dark Energy}},
  \href{https://doi.org/10.1142/S0219887807001928}{\emph{International Journal
  of Geometric Methods in Modern Physics} {\bfseries 04} (2007) 115}
  [\href{https://arxiv.org/abs/hep-th/0601213}{{\ttfamily hep-th/0601213}}].

\bibitem{FR_theory2}
T.P.~{Sotiriou} and V.~{Faraoni}, \emph{{f(R) theories of gravity}},
  \href{https://doi.org/10.1103/RevModPhys.82.451}{\emph{Reviews of Modern
  Physics} {\bfseries 82} (2010) 451}
  [\href{https://arxiv.org/abs/0805.1726}{{\ttfamily 0805.1726}}].

\bibitem{kessence_theory1}
C.~{Armend{\'a}riz-Pic{\'o}n}, T.~{Damour} and V.~{Mukhanov},
  \emph{{k-Inflation}},
  \href{https://doi.org/10.1016/S0370-2693(99)00603-6}{\emph{Phys. Lett. B}
  {\bfseries 458} (1999) 209}
  [\href{https://arxiv.org/abs/hep-th/9904075}{{\ttfamily hep-th/9904075}}].

\bibitem{kessence_theory2}
T.~{Chiba}, T.~{Okabe} and M.~{Yamaguchi}, \emph{{Kinetically driven
  quintessence}}, \href{https://doi.org/10.1103/PhysRevD.62.023511}{\emph{Phys.
  Rev. D} {\bfseries 62} (2000) 023511}
  [\href{https://arxiv.org/abs/astro-ph/9912463}{{\ttfamily
  astro-ph/9912463}}].

\bibitem{Horndeski_theory}
G.W.~{Horndeski}, \emph{{Second-order scalar-tensor field equations in a
  four-dimensional space}},
  \href{https://doi.org/10.1007/BF01807638}{\emph{International Journal of
  Theoretical Physics} {\bfseries 10} (1974) 363}.

\bibitem{GG_theory}
C.~{Deffayet}, S.~{Deser} and G.~{Esposito-Far{\`e}se}, \emph{{Generalized
  Galileons: All scalar models whose curved background extensions maintain
  second-order field equations and stress tensors}},
  \href{https://doi.org/10.1103/PhysRevD.80.064015}{\emph{Phys. Rev. D}
  {\bfseries 80} (2009) 064015}
  [\href{https://arxiv.org/abs/0906.1967}{{\ttfamily 0906.1967}}].

\bibitem{DGP_theory}
G.~{Dvali}, G.~{Gabadadze} and M.~{Porrati}, \emph{{4D gravity on a brane in 5D
  Minkowski space}},
  \href{https://doi.org/10.1016/S0370-2693(00)00669-9}{\emph{Phys. Lett. B}
  {\bfseries 485} (2000) 208}
  [\href{https://arxiv.org/abs/hep-th/0005016}{{\ttfamily hep-th/0005016}}].

\bibitem{probes_review1}
D.~{Huterer} and D.L.~{Shafer}, \emph{{Dark energy two decades after:
  observables, probes, consistency tests}},
  \href{https://doi.org/10.1088/1361-6633/aa997e}{\emph{Reports on Progress in
  Physics} {\bfseries 81} (2018) 016901}
  [\href{https://arxiv.org/abs/1709.01091}{{\ttfamily 1709.01091}}].

\bibitem{probes_review2}
M.~{Moresco}, L.~{Amati}, L.~{Amendola}, S.~{Birrer}, J.P.~{Blakeslee},
  M.~{Cantiello} et~al., \emph{{Unveiling the Universe with emerging
  cosmological probes}},
  \href{https://doi.org/10.1007/s41114-022-00040-z}{\emph{Living Reviews in
  Relativity} {\bfseries 25} (2022) 6}
  [\href{https://arxiv.org/abs/2201.07241}{{\ttfamily 2201.07241}}].

\bibitem{CPL1}
M.~{Chevallier} and D.~{Polarski}, \emph{{Accelerating Universes with Scaling
  Dark Matter}},
  \href{https://doi.org/10.1142/S0218271801000822}{\emph{International Journal
  of Modern Physics D} {\bfseries 10} (2001) 213}
  [\href{https://arxiv.org/abs/gr-qc/0009008}{{\ttfamily gr-qc/0009008}}].

\bibitem{CPL2}
E.V.~{Linder}, \emph{{Exploring the Expansion History of the Universe}},
  \href{https://doi.org/10.1103/PhysRevLett.90.091301}{\emph{Phys. Rev. Lett.}
  {\bfseries 90} (2003) 091301}
  [\href{https://arxiv.org/abs/astro-ph/0208512}{{\ttfamily
  astro-ph/0208512}}].

\bibitem{MG_parametrize1}
R.~{Caldwell}, A.~{Cooray} and A.~{Melchiorri}, \emph{{Constraints on a new
  post-general relativity cosmological parameter}},
  \href{https://doi.org/10.1103/PhysRevD.76.023507}{\emph{Phys. Rev. D}
  {\bfseries 76} (2007) 023507}
  [\href{https://arxiv.org/abs/astro-ph/0703375}{{\ttfamily
  astro-ph/0703375}}].

\bibitem{MG_parametrize2}
L.~{Amendola}, M.~{Kunz} and D.~{Sapone}, \emph{{Measuring the dark side (with
  weak lensing)}},
  \href{https://doi.org/10.1088/1475-7516/2008/04/013}{\emph{JCAP} {\bfseries
  2008} (2008) 013} [\href{https://arxiv.org/abs/0704.2421}{{\ttfamily
  0704.2421}}].

\bibitem{EFT_theory1}
G.~{Gubitosi}, F.~{Piazza} and F.~{Vernizzi}, \emph{{The effective field theory
  of dark energy}},
  \href{https://doi.org/10.1088/1475-7516/2013/02/032}{\emph{JCAP} {\bfseries
  2013} (2013) 032} [\href{https://arxiv.org/abs/1210.0201}{{\ttfamily
  1210.0201}}].

\bibitem{EFT_theory2}
J.~{Bloomfield}, {\'E}.{\'E}.~{Flanagan}, M.~{Park} and S.~{Watson},
  \emph{{Dark energy or modified gravity? An effective field theory approach}},
  \href{https://doi.org/10.1088/1475-7516/2013/08/010}{\emph{JCAP} {\bfseries
  2013} (2013) 010} [\href{https://arxiv.org/abs/1211.7054}{{\ttfamily
  1211.7054}}].

\bibitem{EFT_theory3}
J.~{Gleyzes}, D.~{Langlois}, F.~{Piazza} and F.~{Vernizzi}, \emph{{Essential
  building blocks of dark energy}},
  \href{https://doi.org/10.1088/1475-7516/2013/08/025}{\emph{JCAP} {\bfseries
  2013} (2013) 025} [\href{https://arxiv.org/abs/1304.4840}{{\ttfamily
  1304.4840}}].

\bibitem{EFT_theory4}
N.~{Frusciante}, G.~{Papadomanolakis} and A.~{Silvestri}, \emph{{An extended
  action for the effective field theory of dark energy: a stability analysis
  and a complete guide to the mapping at the basis of EFTCAMB}},
  \href{https://doi.org/10.1088/1475-7516/2016/07/018}{\emph{JCAP} {\bfseries
  2016} (2016) 018} [\href{https://arxiv.org/abs/1601.04064}{{\ttfamily
  1601.04064}}].

\bibitem{EFT_review}
N.~{Frusciante} and L.~{Perenon}, \emph{{Effective field theory of dark energy:
  A review}}, \href{https://doi.org/10.1016/j.physrep.2020.02.004}{\emph{Phys.
  Rep.} {\bfseries 857} (2020) 1}
  [\href{https://arxiv.org/abs/1907.03150}{{\ttfamily 1907.03150}}].

\bibitem{nonlinear_EFT1}
N.~{Frusciante} and G.~{Papadomanolakis}, \emph{{Tackling non-linearities with
  the effective field theory of dark energy and modified gravity}},
  \href{https://doi.org/10.1088/1475-7516/2017/12/014}{\emph{JCAP} {\bfseries
  2017} (2017) 014} [\href{https://arxiv.org/abs/1706.02719}{{\ttfamily
  1706.02719}}].

\bibitem{nonlinear_EFT2}
G.~{Cusin}, M.~{Lewandowski} and F.~{Vernizzi}, \emph{{Nonlinear effective
  theory of dark energy}},
  \href{https://doi.org/10.1088/1475-7516/2018/04/061}{\emph{JCAP} {\bfseries
  2018} (2018) 061} [\href{https://arxiv.org/abs/1712.02782}{{\ttfamily
  1712.02782}}].

\bibitem{quintessence_reconstruct1}
T.~{Chiba} and T.~{Nakamura}, \emph{{Feasibility of reconstructing the
  quintessential potential using SNIa data}},  in \emph{20th Texas Symposium on
  relativistic astrophysics}, vol.~586 of \emph{American Institute of Physics
  Conference Series}, pp.~319--321, AIP, 2001,
  \href{https://doi.org/10.1063/1.1419572}{DOI}.

\bibitem{quintessence_reconstruct2}
C.~{Rubano} and J.D.~{Barrow}, \emph{{Scaling solutions and reconstruction of
  scalar field potentials}},
  \href{https://doi.org/10.1103/PhysRevD.64.127301}{\emph{Phys. Rev. D}
  {\bfseries 64} (2001) 127301}
  [\href{https://arxiv.org/abs/gr-qc/0105037}{{\ttfamily gr-qc/0105037}}].

\bibitem{quintessence_reconstruct3}
B.F.~{Gerke} and G.~{Efstathiou}, \emph{{Probing quintessence: reconstruction
  and parameter estimation from supernovae}},
  \href{https://doi.org/10.1046/j.1365-8711.2002.05612.x}{\emph{Mon. Not. Roy.
  Astron. Soc.} {\bfseries 335} (2002) 33}
  [\href{https://arxiv.org/abs/astro-ph/0201336}{{\ttfamily
  astro-ph/0201336}}].

\bibitem{quintessence_reconstruct4}
X.~{Zhang}, \emph{{Reconstructing holographic quintessence}},
  \href{https://doi.org/10.1016/j.physletb.2007.02.069}{\emph{Phys. Lett. B}
  {\bfseries 648} (2007) 1}
  [\href{https://arxiv.org/abs/astro-ph/0604484}{{\ttfamily
  astro-ph/0604484}}].

\bibitem{quintessence_reconstruct5}
J.-P.~{Wu}, D.-Z.~{Ma} and Y.~{Ling}, \emph{{Quintessence reconstruction of the
  new agegraphic dark energy model}},
  \href{https://doi.org/10.1016/j.physletb.2008.03.071}{\emph{Phys. Lett. B}
  {\bfseries 663} (2008) 152}
  [\href{https://arxiv.org/abs/0805.0546}{{\ttfamily 0805.0546}}].

\bibitem{quintessence_reconstruct6}
A.~{Sangwan}, A.~{Mukherjee} and H.K.~{Jassal}, \emph{{Reconstructing the dark
  energy potential}},
  \href{https://doi.org/10.1088/1475-7516/2018/01/018}{\emph{JCAP} {\bfseries
  2018} (2018) 018} [\href{https://arxiv.org/abs/1712.05143}{{\ttfamily
  1712.05143}}].

\bibitem{quintessence_reconstruct7}
M.~{Park}, M.~{Raveri} and B.~{Jain}, \emph{{Reconstructing quintessence}},
  \href{https://doi.org/10.1103/PhysRevD.103.103530}{\emph{Phys. Rev. D}
  {\bfseries 103} (2021) 103530}
  [\href{https://arxiv.org/abs/2101.04666}{{\ttfamily 2101.04666}}].

\bibitem{STT_reconstruct1}
S.~{Nojiri} and S.D.~{Odintsov}, \emph{{Unifying phantom inflation with
  late-time acceleration: scalar phantom-non-phantom transition model and
  generalized holographic dark energy}},
  \href{https://doi.org/10.1007/s10714-006-0301-6}{\emph{General Relativity and
  Gravitation} {\bfseries 38} (2006) 1285}
  [\href{https://arxiv.org/abs/hep-th/0506212}{{\ttfamily hep-th/0506212}}].

\bibitem{STT_reconstruct2}
S.~{Capozziello}, S.~{Nojiri} and S.D.~{Odintsov}, \emph{{Unified phantom
  cosmology: Inflation, dark energy and dark matter under the same standard}},
  \href{https://doi.org/10.1016/j.physletb.2005.11.012}{\emph{Phys. Lett.s B}
  {\bfseries 632} (2006) 597}
  [\href{https://arxiv.org/abs/hep-th/0507182}{{\ttfamily hep-th/0507182}}].

\bibitem{STT_reconstruct3}
S.~{Capozziello}, S.~{Nojiri} and S.D.~{Odintsov}, \emph{{Dark energy: the
  equation of state description versus scalar-tensor or modified gravity}},
  \href{https://doi.org/10.1016/j.physletb.2006.01.065}{\emph{Phys. Lett. B}
  {\bfseries 634} (2006) 93}
  [\href{https://arxiv.org/abs/hep-th/0512118}{{\ttfamily hep-th/0512118}}].

\bibitem{STT_reconstruct7}
B.~{Boisseau}, G.~{Esposito-Far{\`e}se}, D.~{Polarski} and A.A.~{Starobinsky},
  \emph{{Reconstruction of a Scalar-Tensor Theory of Gravity in an Accelerating
  Universe}}, \href{https://doi.org/10.1103/PhysRevLett.85.2236}{\emph{Phys.
  Rev. Lett.} {\bfseries 85} (2000) 2236}
  [\href{https://arxiv.org/abs/gr-qc/0001066}{{\ttfamily gr-qc/0001066}}].

\bibitem{STT_reconstruct4}
S.~{Nojiri}, S.D.~{Odintsov} and M.~{Sami}, \emph{{Dark energy cosmology from
  higher-order, string-inspired gravity, and its reconstruction}},
  \href{https://doi.org/10.1103/PhysRevD.74.046004}{\emph{Phys. Rev. D}
  {\bfseries 74} (2006) 046004}
  [\href{https://arxiv.org/abs/hep-th/0605039}{{\ttfamily hep-th/0605039}}].

\bibitem{STT_reconstruct5}
G.~{Cognola}, E.~{Elizalde}, S.~{Nojiri}, S.D.~{Odintsov} and S.~{Zerbini},
  \emph{{String-inspired Gauss-Bonnet gravity reconstructed from the universe
  expansion history and yielding the transition from matter dominance to dark
  energy}}, \href{https://doi.org/10.1103/PhysRevD.75.086002}{\emph{Phys. Rev.
  D} {\bfseries 75} (2007) 086002}
  [\href{https://arxiv.org/abs/hep-th/0611198}{{\ttfamily hep-th/0611198}}].

\bibitem{STT_reconstruct6}
S.~{Nojiri} and S.D.~{Odintsov}, \emph{{Modified gravity and its reconstruction
  from the universe expansion history}},  in \emph{Journal of Physics
  Conference Series}, vol.~66 of \emph{Journal of Physics Conference Series},
  p.~012005, IOP, 2007,
  \href{https://doi.org/10.1088/1742-6596/66/1/012005}{DOI}
  [\href{https://arxiv.org/abs/hep-th/0611071}{{\ttfamily hep-th/0611071}}].

\bibitem{STT_reconstruct8}
S.~{Capozziello}, S.~{Nesseris} and L.~{Perivolaropoulos},
  \emph{{Reconstruction of the scalar tensor Lagrangian from a
  {\ensuremath{\Lambda}}CDM background and Noether symmetry}},
  \href{https://doi.org/10.1088/1475-7516/2007/12/009}{\emph{JCAP} {\bfseries
  2007} (2007) 009} [\href{https://arxiv.org/abs/0705.3586}{{\ttfamily
  0705.3586}}].

\bibitem{MG_tomography}
P.~{Brax}, A.-C.~{Davis} and B.~{Li}, \emph{{Modified gravity tomography}},
  \href{https://doi.org/10.1016/j.physletb.2012.08.002}{\emph{Phys. Lett. B}
  {\bfseries 715} (2012) 38} [\href{https://arxiv.org/abs/1111.6613}{{\ttfamily
  1111.6613}}].

\bibitem{FR_reconstruct1}
S.~{Nojiri} and S.D.~{Odintsov}, \emph{{Modified f(R) gravity consistent with
  realistic cosmology: From a matter dominated epoch to a dark energy
  universe}}, \href{https://doi.org/10.1103/PhysRevD.74.086005}{\emph{Phys.
  Rev. D} {\bfseries 74} (2006) 086005}
  [\href{https://arxiv.org/abs/hep-th/0608008}{{\ttfamily hep-th/0608008}}].

\bibitem{FR_reconstruct2}
E.~{Elizalde} and D.~{S{\'a}ez-G{\'o}mez}, \emph{{F(R) cosmology in the
  presence of a phantom fluid and its scalar-tensor counterpart: Towards a
  unified precision model of the evolution of the Universe}},
  \href{https://doi.org/10.1103/PhysRevD.80.044030}{\emph{Phys. Rev. D}
  {\bfseries 80} (2009) 044030}
  [\href{https://arxiv.org/abs/0903.2732}{{\ttfamily 0903.2732}}].

\bibitem{FR_reconstruct3}
K.~{Karami} and M.S.~{Khaledian}, \emph{{Reconstructing f( R) modified gravity
  from ordinary and entropy-corrected versions of the holographic and new
  agegraphic dark energy models}},
  \href{https://doi.org/10.1007/JHEP03(2011)086}{\emph{Journal of High Energy
  Physics} {\bfseries 2011} (2011) 86}
  [\href{https://arxiv.org/abs/1004.1805}{{\ttfamily 1004.1805}}].

\bibitem{FR_reconstruct4}
S.~{Carloni}, R.~{Goswami} and P.K.S.~{Dunsby}, \emph{{A new approach to
  reconstruction methods in f(R) gravity}},
  \href{https://doi.org/10.1088/0264-9381/29/13/135012}{\emph{Classical and
  Quantum Gravity} {\bfseries 29} (2012) 135012}
  [\href{https://arxiv.org/abs/1005.1840}{{\ttfamily 1005.1840}}].

\bibitem{FR_reconstruct5}
M.J.S.~{Houndjo}, \emph{{Reconstruction of f(R, t) Gravity Describing Matter
  Dominated and Accelerated Phases}},
  \href{https://doi.org/10.1142/S0218271812500034}{\emph{International Journal
  of Modern Physics D} {\bfseries 21} (2012) 1250003}
  [\href{https://arxiv.org/abs/1107.3887}{{\ttfamily 1107.3887}}].

\bibitem{FR_reconstruct6}
M.~{Jamil}, D.~{Momeni}, M.~{Raza} and R.~{Myrzakulov}, \emph{{Reconstruction
  of some cosmological models in f( R, T) cosmology}},
  \href{https://doi.org/10.1140/epjc/s10052-012-1999-9}{\emph{European Physical
  Journal C} {\bfseries 72} (2012) 1999}
  [\href{https://arxiv.org/abs/1107.5807}{{\ttfamily 1107.5807}}].

\bibitem{FR_reconstruct7}
C.P.~{Singh} and V.~{Singh}, \emph{{Reconstruction of modified gravity with
  perfect fluid cosmological models}},
  \href{https://doi.org/10.1007/s10714-014-1696-0}{\emph{General Relativity and
  Gravitation} {\bfseries 46} (2014) 1696}.

\bibitem{FR_reconstruct8}
E.~{Elizalde}, R.~{Myrzakulov}, V.V.~{Obukhov} and D.~{S{\'a}ez-G{\'o}mez},
  \emph{{{\ensuremath{\Lambda}}CDM epoch reconstruction from F(R, G) and
  modified Gauss-Bonnet gravities}},
  \href{https://doi.org/10.1088/0264-9381/27/9/095007}{\emph{Classical and
  Quantum Gravity} {\bfseries 27} (2010) 095007}
  [\href{https://arxiv.org/abs/1001.3636}{{\ttfamily 1001.3636}}].

\bibitem{FR_reconstruct9}
M.~{Hamani Daouda}, M.E.~{Rodrigues} and M.J.S.~{Houndjo},
  \emph{{Reconstruction of f( T) gravity according to holographic dark
  energy}},
  \href{https://doi.org/10.1140/epjc/s10052-012-1893-5}{\emph{European Physical
  Journal C} {\bfseries 72} (2012) 1893}
  [\href{https://arxiv.org/abs/1111.6575}{{\ttfamily 1111.6575}}].

\bibitem{FR_reconstruct10}
E.A.~{Elkhateeb}, \emph{{Reconstruction of f(R) Gravity from Cosmological
  Unified Dark Fluid Model}},
  \href{https://doi.org/10.1007/s10701-023-00751-5}{\emph{Foundations of
  Physics} {\bfseries 54} (2024) 18}
  [\href{https://arxiv.org/abs/2301.13858}{{\ttfamily 2301.13858}}].

\bibitem{kessence_reconstruct1}
S.~{Tsujikawa}, \emph{{Reconstruction of general scalar-field dark energy
  models}}, \href{https://doi.org/10.1103/PhysRevD.72.083512}{\emph{Phys. Rev.
  D} {\bfseries 72} (2005) 083512}
  [\href{https://arxiv.org/abs/astro-ph/0508542}{{\ttfamily
  astro-ph/0508542}}].

\bibitem{kessence_reconstruct2}
A.A.~{Sen}, \emph{{Reconstructing k-essence}},
  \href{https://doi.org/10.1088/1475-7516/2006/03/010}{\emph{JCAP} {\bfseries
  2006} (2006) 010} [\href{https://arxiv.org/abs/astro-ph/0512406}{{\ttfamily
  astro-ph/0512406}}].

\bibitem{kessence_reconstruct3}
J.~{Matsumoto} and S.~{Nojiri}, \emph{{Reconstruction of k-essence model}},
  \href{https://doi.org/10.1016/j.physletb.2010.03.030}{\emph{Phys. Lett. B}
  {\bfseries 687} (2010) 236}
  [\href{https://arxiv.org/abs/1001.0220}{{\ttfamily 1001.0220}}].

\bibitem{kessence_reconstruct4}
D.~{Perkovi{\'c}} and H.~{{\v{S}}tefan{\v{c}}i{\'c}}, \emph{{Analytical
  reconstruction of equivalent purely kinetic k-essence description for [ image
  ] barotropic fluid models}},
  \href{https://doi.org/10.1088/1361-6382/adbeb3}{\emph{Classical and Quantum
  Gravity} {\bfseries 42} (2025) 075017}.

\bibitem{ghost_condensation_reconstruct1}
X.~{Zhang}, \emph{{Dynamical vacuum energy, holographic quintom, and the
  reconstruction of scalar-field dark energy}},
  \href{https://doi.org/10.1103/PhysRevD.74.103505}{\emph{Phys. Rev. D}
  {\bfseries 74} (2006) 103505}
  [\href{https://arxiv.org/abs/astro-ph/0609699}{{\ttfamily
  astro-ph/0609699}}].

\bibitem{ghost_condensation_reconstruct2}
J.~{Zhang}, X.~{Zhang} and H.~{Liu}, \emph{{Reconstructing Generalized Ghost
  Condensate Model with Dynamical Dark Energy Parametrizations and
  Observational Datasets}},
  \href{https://doi.org/10.1142/S0217732308023505}{\emph{Modern Physics Letters
  A} {\bfseries 23} (2008) 139}
  [\href{https://arxiv.org/abs/astro-ph/0612642}{{\ttfamily
  astro-ph/0612642}}].

\bibitem{cG_reconstruct1}
A.~{Ijjas} and P.J.~{Steinhardt}, \emph{{Classically Stable Nonsingular
  Cosmological Bounces}},
  \href{https://doi.org/10.1103/PhysRevLett.117.121304}{\emph{Phys. Rev. Lett.}
  {\bfseries 117} (2016) 121304}
  [\href{https://arxiv.org/abs/1606.08880}{{\ttfamily 1606.08880}}].

\bibitem{cG_reconstruct2}
D.A.~{Dobre}, A.V.~{Frolov}, J.T.~{G{\'a}lvez Ghersi}, S.~{Ramazanov} and
  A.~{Vikman}, \emph{{Unbraiding the bounce: superluminality around the
  corner}}, \href{https://doi.org/10.1088/1475-7516/2018/03/020}{\emph{JCAP}
  {\bfseries 2018} (2018) 020}
  [\href{https://arxiv.org/abs/1712.10272}{{\ttfamily 1712.10272}}].

\bibitem{cG_reconstruct3}
R.~{Arjona}, W.~{Cardona} and S.~{Nesseris}, \emph{{Designing Horndeski and the
  effective fluid approach}},
  \href{https://doi.org/10.1103/PhysRevD.100.063526}{\emph{Phys. Rev. D}
  {\bfseries 100} (2019) 063526}
  [\href{https://arxiv.org/abs/1904.06294}{{\ttfamily 1904.06294}}].

\bibitem{cG_reconstruct4}
R.~{Arjona}, \emph{{The effective fluid approach for modified gravity}},
  \href{https://doi.org/10.48550/arXiv.2010.04764}{\emph{arXiv e-prints} (2020)
  arXiv:2010.04764} [\href{https://arxiv.org/abs/2010.04764}{{\ttfamily
  2010.04764}}].

\bibitem{cG_reconstruct5}
R.C.~{Bernardo} and J.~{Levi Said}, \emph{{A data-driven reconstruction of
  Horndeski gravity via the Gaussian processes}},
  \href{https://doi.org/10.1088/1475-7516/2021/09/014}{\emph{JCAP} {\bfseries
  2021} (2021) 014} [\href{https://arxiv.org/abs/2105.12970}{{\ttfamily
  2105.12970}}].

\bibitem{cG_reconstruct6}
S.~{Nesseris}, \emph{{The Effective Fluid Approach for Modified Gravity and Its
  Applications}},
  \href{https://doi.org/10.3390/universe9010013}{\emph{Universe} {\bfseries 9}
  (2022) 13} [\href{https://arxiv.org/abs/2212.12768}{{\ttfamily 2212.12768}}].

\bibitem{cG_reconstruct7}
R.C.~{Bernardo} and I.~{Vega}, \emph{{Tailoring cosmologies in cubic
  shift-symmetric Horndeski gravity}},
  \href{https://doi.org/10.1088/1475-7516/2019/10/058}{\emph{JCAP} {\bfseries
  2019} (2019) 058} [\href{https://arxiv.org/abs/1903.12578}{{\ttfamily
  1903.12578}}].

\bibitem{k_inflation1}
R.~{Herrera}, \emph{{Reconstructing k -essence: Unifying the attractor
  n$_{S}$(N ) and the swampland criteria}},
  \href{https://doi.org/10.1103/PhysRevD.102.123508}{\emph{Phys. Rev. D}
  {\bfseries 102} (2020) 123508}
  [\href{https://arxiv.org/abs/2009.01355}{{\ttfamily 2009.01355}}].

\bibitem{k_inflation2}
L.~{Sebastiani}, S.~{Myrzakul} and R.~{Myrzakulov}, \emph{{Reconstruction of
  k-essence inflation in Horndeski gravity}},
  \href{https://doi.org/10.1140/epjp/i2017-11695-1}{\emph{European Physical
  Journal Plus} {\bfseries 132} (2017) 433}
  [\href{https://arxiv.org/abs/1702.00064}{{\ttfamily 1702.00064}}].

\bibitem{k_inflation3}
R.~{Herrera}, M.~{Housset}, C.~{Osses} and N.~{Videla}, \emph{{Reconstructing
  k-inflation from n$_{s}$(N) and reheating constraints}},
  \href{https://doi.org/10.1016/j.dark.2023.101386}{\emph{Physics of the Dark
  Universe} {\bfseries 43} (2024) 101386}
  [\href{https://arxiv.org/abs/2305.05042}{{\ttfamily 2305.05042}}].

\bibitem{G_inflation1}
R.~{Herrera}, \emph{{Reconstructing G inflation: From the attractors n$_{S}$(N
  ) and r (N )}}, \href{https://doi.org/10.1103/PhysRevD.98.023542}{\emph{Phys.
  Rev. D} {\bfseries 98} (2018) 023542}
  [\href{https://arxiv.org/abs/1805.01007}{{\ttfamily 1805.01007}}].

\bibitem{DESI:2025zgx}
{\scshape DESI} collaboration, \emph{{DESI DR2 Results II: Measurements of
  Baryon Acoustic Oscillations and Cosmological Constraints}},
  \href{https://arxiv.org/abs/2503.14738}{{\ttfamily 2503.14738}}.

\bibitem{EFT_reconstruct1}
J.~{Kennedy}, L.~{Lombriser} and A.~{Taylor}, \emph{{Reconstructing Horndeski
  models from the effective field theory of dark energy}},
  \href{https://doi.org/10.1103/PhysRevD.96.084051}{\emph{Phys. Rev. D}
  {\bfseries 96} (2017) 084051}.

\bibitem{EFT_reconstruct2}
J.~{Kennedy}, L.~{Lombriser} and A.~{Taylor}, \emph{{Reconstructing Horndeski
  theories from phenomenological modified gravity and dark energy models on
  cosmological scales}},
  \href{https://doi.org/10.1103/PhysRevD.98.044051}{\emph{Phys. Rev. D}
  {\bfseries 98} (2018) 044051}
  [\href{https://arxiv.org/abs/1804.04582}{{\ttfamily 1804.04582}}].

\bibitem{EFT_reconstruct3}
J.~{Kennedy}, L.~{Lombriser} and A.~{Taylor}, \emph{{Screening and degenerate
  kinetic self-acceleration from the nonlinear freedom of reconstructed
  Horndeski theories}},
  \href{https://doi.org/10.1103/PhysRevD.100.044034}{\emph{Phys. Rev. D}
  {\bfseries 100} (2019) 044034}
  [\href{https://arxiv.org/abs/1902.09853}{{\ttfamily 1902.09853}}].

\bibitem{EFT_reconstruct4}
C.~{Renevey}, J.~{Kennedy} and L.~{Lombriser}, \emph{{Parameterised
  post-Newtonian formalism for the effective field theory of dark energy via
  screened reconstructed Horndeski theories}},
  \href{https://doi.org/10.1088/1475-7516/2020/12/032}{\emph{JCAP} {\bfseries
  2020} (2020) 032} [\href{https://arxiv.org/abs/2006.09910}{{\ttfamily
  2006.09910}}].

\bibitem{Horndeski_review}
T.~{Kobayashi}, \emph{{Horndeski theory and beyond: a review}},
  \href{https://doi.org/10.1088/1361-6633/ab2429}{\emph{Reports on Progress in
  Physics} {\bfseries 82} (2019) 086901}
  [\href{https://arxiv.org/abs/1901.07183}{{\ttfamily 1901.07183}}].

\bibitem{LIGOScientific:2017vwq}
{\scshape LIGO Scientific, Virgo} collaboration, \emph{{GW170817: Observation
  of Gravitational Waves from a Binary Neutron Star Inspiral}},
  \href{https://doi.org/10.1103/PhysRevLett.119.161101}{\emph{Phys. Rev. Lett.}
  {\bfseries 119} (2017) 161101}
  [\href{https://arxiv.org/abs/1710.05832}{{\ttfamily 1710.05832}}].

\bibitem{GW17_1}
P.~{Creminelli} and F.~{Vernizzi}, \emph{{Dark Energy after GW170817 and
  GRB170817A}},
  \href{https://doi.org/10.1103/PhysRevLett.119.251302}{\emph{Phys. Rev. Lett.}
  {\bfseries 119} (2017) 251302}.

\bibitem{GW17_2}
J.~{Sakstein} and B.~{Jain}, \emph{{Implications of the Neutron Star Merger
  GW170817 for Cosmological Scalar-Tensor Theories}},
  \href{https://doi.org/10.1103/PhysRevLett.119.251303}{\emph{Phys. Rev. Lett.}
  {\bfseries 119} (2017) 251303}
  [\href{https://arxiv.org/abs/1710.05893}{{\ttfamily 1710.05893}}].

\bibitem{GW17_3}
J.M.~{Ezquiaga} and M.~{Zumalac{\'a}rregui}, \emph{{Dark Energy After GW170817:
  Dead Ends and the Road Ahead}},
  \href{https://doi.org/10.1103/PhysRevLett.119.251304}{\emph{Phys. Rev. Lett.}
  {\bfseries 119} (2017) 251304}
  [\href{https://arxiv.org/abs/1710.05901}{{\ttfamily 1710.05901}}].

\bibitem{GW17_4}
T.~{Baker}, E.~{Bellini}, P.G.~{Ferreira}, M.~{Lagos}, J.~{Noller} and
  I.~{Sawicki}, \emph{{Strong Constraints on Cosmological Gravity from GW170817
  and GRB 170817A}},
  \href{https://doi.org/10.1103/PhysRevLett.119.251301}{\emph{Phys. Rev. Lett.}
  {\bfseries 119} (2017) 251301}
  [\href{https://arxiv.org/abs/1710.06394}{{\ttfamily 1710.06394}}].

\bibitem{alpha_basis}
E.~{Bellini} and I.~{Sawicki}, \emph{{Maximal freedom at minimum cost: linear
  large-scale structure in general modifications of gravity}},
  \href{https://doi.org/10.1088/1475-7516/2014/07/050}{\emph{JCAP} {\bfseries
  2014} (2014) 050} [\href{https://arxiv.org/abs/1404.3713}{{\ttfamily
  1404.3713}}].

\bibitem{EFTCAMB1}
B.~{Hu}, M.~{Raveri}, N.~{Frusciante} and A.~{Silvestri}, \emph{{Effective
  field theory of cosmic acceleration: An implementation in CAMB}},
  \href{https://doi.org/10.1103/PhysRevD.89.103530}{\emph{Phys. Rev. D}
  {\bfseries 89} (2014) 103530}
  [\href{https://arxiv.org/abs/1312.5742}{{\ttfamily 1312.5742}}].

\bibitem{EFTCAMB2}
B.~{Hu}, M.~{Raveri}, N.~{Frusciante} and A.~{Silvestri},
  \emph{{EFTCAMB/EFTCosmoMC: Numerical Notes v3.0}},
  \href{https://doi.org/10.48550/arXiv.1405.3590}{\emph{arXiv e-prints} (2014)
  arXiv:1405.3590} [\href{https://arxiv.org/abs/1405.3590}{{\ttfamily
  1405.3590}}].

\bibitem{hiclass1}
M.~{Zumalac{\'a}rregui}, E.~{Bellini}, I.~{Sawicki}, J.~{Lesgourgues} and
  P.G.~{Ferreira}, \emph{{hi\_class: Horndeski in the Cosmic Linear Anisotropy
  Solving System}},
  \href{https://doi.org/10.1088/1475-7516/2017/08/019}{\emph{JCAP} {\bfseries
  2017} (2017) 019} [\href{https://arxiv.org/abs/1605.06102}{{\ttfamily
  1605.06102}}].

\bibitem{hiclass2}
E.~{Bellini}, I.~{Sawicki} and M.~{Zumalac{\'a}rregui}, \emph{{hi\_class
  background evolution, initial conditions and approximation schemes}},
  \href{https://doi.org/10.1088/1475-7516/2020/02/008}{\emph{JCAP} {\bfseries
  2020} (2020) 008} [\href{https://arxiv.org/abs/1909.01828}{{\ttfamily
  1909.01828}}].

\bibitem{STT_phantom_crossing1}
S.~{Nojiri}, S.D.~{Odintsov} and S.~{Tsujikawa}, \emph{{Properties of
  singularities in the (phantom) dark energy universe}},
  \href{https://doi.org/10.1103/PhysRevD.71.063004}{\emph{Phys. Rev. D}
  {\bfseries 71} (2005) 063004}
  [\href{https://arxiv.org/abs/hep-th/0501025}{{\ttfamily hep-th/0501025}}].

\bibitem{STT_phantom_crossing2}
T.~{Clemson}, K.~{Koyama}, G.-B.~{Zhao}, R.~{Maartens} and J.~{V{\"a}liviita},
  \emph{{Interacting dark energy: Constraints and degeneracies}},
  \href{https://doi.org/10.1103/PhysRevD.85.043007}{\emph{Phys. Rev. D}
  {\bfseries 85} (2012) 043007}
  [\href{https://arxiv.org/abs/1109.6234}{{\ttfamily 1109.6234}}].

\bibitem{k_essence_stability}
A.~{Vikman}, \emph{{Can dark energy evolve to the phantom?}},
  \href{https://doi.org/10.1103/PhysRevD.71.023515}{\emph{Phys. Rev. D}
  {\bfseries 71} (2005) 023515}
  [\href{https://arxiv.org/abs/astro-ph/0407107}{{\ttfamily
  astro-ph/0407107}}].

\bibitem{k_essence_stability2}
R.R.~{Caldwell} and M.~{Doran}, \emph{{Dark-energy evolution across the
  cosmological-constant boundary}},
  \href{https://doi.org/10.1103/PhysRevD.72.043527}{\emph{Phys. Rev. D}
  {\bfseries 72} (2005) 043527}
  [\href{https://arxiv.org/abs/astro-ph/0501104}{{\ttfamily
  astro-ph/0501104}}].

\bibitem{k_essence_stability3}
J.-Q.~{Xia}, Y.-F.~{Cai}, T.-T.~{Qiu}, G.-B.~{Zhao} and X.~{Zhang},
  \emph{{Constraints on the Sound Speed of Dynamical Dark Energy}},
  \href{https://doi.org/10.1142/S0218271808012784}{\emph{International Journal
  of Modern Physics D} {\bfseries 17} (2008) 1229}
  [\href{https://arxiv.org/abs/astro-ph/0703202}{{\ttfamily
  astro-ph/0703202}}].

\bibitem{k_essence_stability4}
Y.-F.~{Cai}, E.N.~{Saridakis}, M.R.~{Setare} and J.-Q.~{Xia}, \emph{{Quintom
  cosmology: Theoretical implications and observations}},
  \href{https://doi.org/10.1016/j.physrep.2010.04.001}{\emph{Phys. Rep.}
  {\bfseries 493} (2010) 1} [\href{https://arxiv.org/abs/0909.2776}{{\ttfamily
  0909.2776}}].

\bibitem{factorizable_k_essence_stability}
L.R.~{Abramo} and N.~{Pinto-Neto}, \emph{{Stability of phantom k-essence
  theories}}, \href{https://doi.org/10.1103/PhysRevD.73.063522}{\emph{Phys.
  Rev. D} {\bfseries 73} (2006) 063522}
  [\href{https://arxiv.org/abs/astro-ph/0511562}{{\ttfamily
  astro-ph/0511562}}].

\bibitem{cubic_Galileon1}
A.~{Nicolis}, R.~{Rattazzi} and E.~{Trincherini}, \emph{{Galileon as a local
  modification of gravity}},
  \href{https://doi.org/10.1103/PhysRevD.79.064036}{\emph{Phys. Rev. D}
  {\bfseries 79} (2009) 064036}
  [\href{https://arxiv.org/abs/0811.2197}{{\ttfamily 0811.2197}}].

\bibitem{cubic_Galileon2}
C.~{Deffayet}, G.~{Esposito-Far{\`e}se} and A.~{Vikman}, \emph{{Covariant
  Galileon}}, \href{https://doi.org/10.1103/PhysRevD.79.084003}{\emph{Phys.
  Rev. D} {\bfseries 79} (2009) 084003}
  [\href{https://arxiv.org/abs/0901.1314}{{\ttfamily 0901.1314}}].

\bibitem{tracker_linear}
A.~{de Felice} and S.~{Tsujikawa}, \emph{{Cosmology of a Covariant Galileon
  Field}}, \href{https://doi.org/10.1103/PhysRevLett.105.111301}{\emph{Phys.
  Rev. Lett.} {\bfseries 105} (2010) 111301}
  [\href{https://arxiv.org/abs/1007.2700}{{\ttfamily 1007.2700}}].

\bibitem{tracker_power}
A.~{De Felice} and S.~{Tsujikawa}, \emph{{Conditions for the cosmological
  viability of the most general scalar-tensor theories and their applications
  to extended Galileon dark energy models}},
  \href{https://doi.org/10.1088/1475-7516/2012/02/007}{\emph{JCAP} {\bfseries
  2012} (2012) 007} [\href{https://arxiv.org/abs/1110.3878}{{\ttfamily
  1110.3878}}].

\bibitem{GCCG1}
F.~{Giacomello}, A.~{De Felice} and S.~{Ansoldi}, \emph{{Bounds from ISW-galaxy
  cross-correlations on generalized covariant Galileon models}},
  \href{https://doi.org/10.1088/1475-7516/2019/03/038}{\emph{JCAP} {\bfseries
  2019} (2019) 038} [\href{https://arxiv.org/abs/1811.10885}{{\ttfamily
  1811.10885}}].

\bibitem{GCCG2}
N.~{Frusciante}, S.~{Peirone}, L.~{Atayde} and A.~{De Felice},
  \emph{{Phenomenology of the generalized cubic covariant Galileon model and
  cosmological bounds}},
  \href{https://doi.org/10.1103/PhysRevD.101.064001}{\emph{Phys. Rev. D}
  {\bfseries 101} (2020) 064001}
  [\href{https://arxiv.org/abs/1912.07586}{{\ttfamily 1912.07586}}].

\bibitem{GGC1}
R.~{Kase} and S.~{Tsujikawa}, \emph{{Dark energy scenario consistent with
  GW170817 in theories beyond Horndeski gravity}},
  \href{https://doi.org/10.1103/PhysRevD.97.103501}{\emph{Phys. Rev. D}
  {\bfseries 97} (2018) 103501}
  [\href{https://arxiv.org/abs/1802.02728}{{\ttfamily 1802.02728}}].

\bibitem{GGC2}
S.~{Peirone}, G.~{Benevento}, N.~{Frusciante} and S.~{Tsujikawa},
  \emph{{Cosmological data favor Galileon ghost condensate over
  {\ensuremath{\Lambda}} CDM}},
  \href{https://doi.org/10.1103/PhysRevD.100.063540}{\emph{Phys. Rev. D}
  {\bfseries 100} (2019) 063540}
  [\href{https://arxiv.org/abs/1905.05166}{{\ttfamily 1905.05166}}].

\bibitem{kinetic_braiding_theory}
C.~{Deffayet}, O.~{Pujol{\`a}s}, I.~{Sawicki} and A.~{Vikman}, \emph{{Imperfect
  dark energy from kinetic gravity braiding}},
  \href{https://doi.org/10.1088/1475-7516/2010/10/026}{\emph{JCAP} {\bfseries
  2010} (2010) 026} [\href{https://arxiv.org/abs/1008.0048}{{\ttfamily
  1008.0048}}].

\bibitem{tracker_analysis}
A.~{Barreira}, B.~{Li}, C.M.~{Baugh} and S.~{Pascoli}, \emph{{The observational
  status of Galileon gravity after Planck}},
  \href{https://doi.org/10.1088/1475-7516/2014/08/059}{\emph{JCAP} {\bfseries
  2014} (2014) 059} [\href{https://arxiv.org/abs/1406.0485}{{\ttfamily
  1406.0485}}].

\bibitem{MG_tomography2}
P.~Brax, A.-C.~Davis, B.~Li and H.A.~Winther, \emph{{A Unified Description of
  Screened Modified Gravity}},
  \href{https://doi.org/10.1103/PhysRevD.86.044015}{\emph{Phys. Rev. D}
  {\bfseries 86} (2012) 044015}
  [\href{https://arxiv.org/abs/1203.4812}{{\ttfamily 1203.4812}}].

\bibitem{MG_tomography_sim1}
P.~{Brax}, A.-C.~{Davis}, B.~{Li}, H.A.~{Winther} and G.-B.~{Zhao},
  \emph{{Systematic simulations of modified gravity: symmetron and dilaton
  models}}, \href{https://doi.org/10.1088/1475-7516/2012/10/002}{\emph{JCAP}
  {\bfseries 2012} (2012) 002}
  [\href{https://arxiv.org/abs/1206.3568}{{\ttfamily 1206.3568}}].

\bibitem{MG_tomography_sim2}
P.~Brax, A.-C.~Davis, B.~Li, H.A.~Winther and G.-B.~Zhao, \emph{{Systematic
  simulations of modified gravity: chameleon models}},
  \href{https://doi.org/10.1088/1475-7516/2013/04/029}{\emph{JCAP} {\bfseries
  04} (2013) 029} [\href{https://arxiv.org/abs/1303.0007}{{\ttfamily
  1303.0007}}].

\end{thebibliography}\endgroup

\end{document}